\documentclass{article}
\usepackage{jheppub}
\usepackage{amsmath}
\usepackage{amssymb}
\usepackage{amsfonts} 
\usepackage{graphbox}
\numberwithin{equation}{section}
\usepackage{subfig}
\usepackage{imakeidx}
\usepackage{tabularx}
\usepackage{adjustbox}
\usepackage{float}

\begin{document}

\title{After the Fluid: Subexponential Decay in AdS$_4$}
\author{John~R.~V.~Crump and}
\author{Jorge~E.~Santos,}

\affiliation{DAMTP, Centre for Mathematical Sciences, University of Cambridge, Wilberforce Road, Cambridge CB3 0WA, UK}
\emailAdd{jrvc2@cam.ac.uk}
\emailAdd{jss55@cam.ac.uk}

\abstract{
We study the late-time behaviour of nonlinear perturbations of Schwarzschild-AdS$_4$ black branes and show that real-analytic initial data generically enter a regime controlled by the large-$k$ tail of the quasinormal mode spectrum $\{\omega_{k\,n}\}$. Using the asymptotic scaling $\mathrm{Im}\,\omega_{k\,n} \sim k^{-1/5}$ of the planar AdS$_4$ black brane, we derive a universal prediction that boundary observables decay in a stretched-exponential manner, specifically as $\exp(-c\, t^{5/6})$ up to a mild polynomial prefactor. Fully nonlinear numerical evolutions employing Fourier spectral and discontinuous Galerkin methods confirm this behaviour for small black holes and show consistent scaling - after suppressing long-lived low-$k$ modes - for larger ones. These results indicate that stretched-exponential decay with exponent $5/6$ is a robust feature of AdS$_4$ gravitational dynamics with real-analytic data, arising from geometric-optics physics rather than hydrodynamic modes.}

\maketitle

\section{Introduction}

The AdS/CFT correspondence provides a startling duality between gravity and quantum field theory \cite{Maldacena:1997re, Aharony:1999ti, Witten:1998qj}. Aside from being a tantalising clue towards a unified theory of quantum gravity, it has also become one of the most powerful tools at our disposal for exploring otherwise intractable physics - such as the behaviour of quantum field theories at strong coupling. One of the most well-studied realisations of AdS/CFT is the duality between a bulk AdS$_4$ spacetime containing a black hole with planar horizon topology and a boundary CFT (in the limit of large $N$ and large coupling) on a fixed $\mathbb{R}^{1,2}$ background at finite temperature, with the Hawking temperature of the black hole equalling the temperature of the boundary theory. Remarkably, long-wavelength low-frequency bulk physics manifests itself as \textit{hydrodynamics} on the boundary, being described by a relativistic generalisation of the Navier stokes equations of fluid dynamics fame -  a result known as the fluid/gravity correspondence \cite{Bhattacharyya:2007vjd, Hubeny:2011hd}. This has proven to be an exciting arena for studying fluid dynamics in its own right, where the described fluid resembles a hot viscous quark-gluon plasma, with gravitational calculations yielding the fluid transport coefficients of this plasma among other seminal results \cite{Kovtun:2004de, Policastro:2002se}. 

This is a natural setting for the use of numerical methods, which have a long and storied history of applications to the study of black holes. By performing numerical time evolutions of perturbed AdS$_4$ planar black hole spacetimes using Einstein's equations, you are also simultaneously doing numerical fluid dynamics simulations on the boundary for free. A planar black hole is infinite in spatial extent, so to make the system more amenable to numerics it is a common choice to impose periodicity in the planar directions \cite{Adams:2013vsa, Balasubramanian_2014, Westernacher-Schneider:2017xie}. For planar coordinates $(x,y)$ on $\mathbb{R}^2$, this is done by imposing $x\sim x+L_{x}$ and $y \sim y+L_y$ for circumferences $L_x$ and $L_y$ - in doing so, the horizon topology becomes \textit{toroidal}. More generally, we could let the horizon be a tilted torus with modular parameter $\tau$ of our choosing - in this paper we consider only purely imaginary $\tau$ for simplicity. Toroidal Schwarzschild-AdS black hole spacetimes have been studied extensively in the literature \cite{Dunn_2016, Dunn:2018let, Dunn:2018xdm, Horowitz:2020tpa}, with the nonlinear stability of the Einstein-Klein-Gordon system established in \cite{Dunn:2018xdm} for initial data that does not break translational invariance along the toroidal directions but allows the ratio of the periods of the torus to vary with time. Furthermore, \cite{Dunn_2016, Dunn:2018let} established \textit{polynomial} decay for the Klein-Gordon equation on a fixed toroidal Schwarzschild-AdS background for initial data in a certain Sobolev space and for an open set of Klein-Gordon masses above the Breitenlohner-Freedman bound. This result arises due to the existence of null geodesics that remain outside the horizon for arbitrarily long times.

At late times, small perturbations of the black hole can be approximated by superpositions of \textit{quasinormal modes} (QNMs) - finite-energy solutions to the linearised Einstein equations around the black brane background, labelled by a wavenumber $k$ and an overtone number $n$. Periodicity along the $x$ and $y$ directions relates the allowed wavenumbers $k$ to integer multiples of $2\pi/L_x$ and $2\pi/L_y$, respectively. Although the QNM spectrum does not form a complete basis for the full solution space \cite{Warnick:2013hba, Warnick:2022hnc}, it is nevertheless expected to capture the essential behaviour in the asymptotic late-time regime. Indeed, \cite{Warnick:2013hba} shows that, for generic initial data, the best one can obtain is an \textit{asymptotic} expansion valid at large times. At small wavenumbers, the quasinormal mode frequencies with $n=0$ have a $k$-dependence that scales with $-iD\,k^2$ - precisely the scaling expected for a diffusive hydrodynamic shear mode with diffusion constant $D$. This leads to the behaviour observed in many studies done of this system in the long-wavelength regime - diffusion smoothing out the fluid back to equilibrium, with the lowest wavenumber modes being those that survive the longest.

However, this is not the end of the story. Modes with $n=0$ do \textit{not} look like a diffusive hydrodynamic modes for all $k$. As $k \to \infty$, the decay rate of the modes follows the asymptotic scaling $\sim k^{-1/5}$ \cite{Festuccia:2008zx}. This behaviour holds not only for $n=0$ but, in fact, for all fixed overtone numbers $n$. The modes become \textit{longer} lived with $k$ rather than shorter. The presence of such modes is intimately related to the aforementioned existence of null geodesics that remain in the black hole exterior for arbitrarily long times. In the CFT, these modes appear as a sequence of poles of the retarded Green's function approaching the lightcone \cite{Festuccia:2008zx}. If the system was just hydrodynamics then the late time behaviour would be dominated by the longest lived low-wavenumber mode with $k=2\pi/L$. But the system is \textit{not} just hydrodynamics, and there are infinitely many modes at large $k$ with slower decay rates than the $k=2\pi/L$ mode. Assuming the linearised picture accurately portrays the late time behaviour, this leads us to predict a regime at very late times when the $k=2\pi/L$ mode decays to be smaller than the next mode with an even smaller decay rate - typically a mode with very large $k$. In this regime, the character of the decay of the perturbation will be dictated by the long spectral tail of the perturbation and the asymptotic $k^{-1/5}$ scaling of the QNM decay rates. We make an explicit prediction for this behaviour in the case of real-analytic initial data where we predict the decay to be subexponential with time. Furthermore, we expect the perturbation to become increasingly spatially localised with time due to a broadening of the spectrum, which is in stark contrast to the diffusion seen in the hydrodynamics regime. For small black holes, \textit{i.e.} those with small $L_i$ where the lowest nonzero wavenumber $k_i=2\pi/L_i$ is sufficiently large, we expect this to be the \textit{only} regime, due to the absence of the long-wavelength hydrodynamic sector in their QNM spectra. We then present two fully nonlinear numerical time evolutions of perturbed AdS$_4$ toroidal black hole spacetimes to support these predictions.

This paper is organised as follows. Section 2 sets up the gravitational system in detail: we describe the Schwarzschild–AdS$_4$ black brane geometry, present the time‑dependent metric ansatz, outline the nested Bondi–Sachs evolution scheme, and derive the analytic prediction for the late‑time behaviour governed by the large‑$k$ quasinormal‑mode tail. Section 3 contains our fully nonlinear numerical results, beginning with small‑$L$ black holes that lack a hydrodynamic sector and then moving to larger black holes where long‑wavelength modes coexist with the high‑$k$ tail. In Section 4, we interpret the emerging dynamics in terms of void formation and Knudsen‑like transport on the boundary, clarifying the physical mechanism behind the stretched‑exponential decay. Finally, Section 5 discusses the implications of our findings, outlines limitations of the present numerical approach, and highlights directions for future work.

\section{Setup of the problem}

\subsection{The Schwarzschild-AdS$_4$ black brane}
The Schwarzschild-AdS$_4$ black brane, in Schwarzschild coordinates $(T,r,X,Y)$, is given by
\begin{subequations}
\begin{equation}
\mathrm{d}s^2
 = -f(r)\,\mathrm{d}T^2
   + \frac{\mathrm{d}r^2}{f(r)}
   + r^2\bigl(\mathrm{d}X^2 + \mathrm{d}Y^2\bigr)\,,
\end{equation}
with
\begin{equation}
f(r) = \frac{r^2}{\ell^2} - \frac{r_h^{3}}{r\,\ell^2}\,.
\end{equation}
\end{subequations}
We take $T\in\mathbb{R}$, and impose periodic identifications
$(X,Y)\sim (X+L_X,\,Y+L_Y)$.

Although the metric appears singular at $r=r_h$, this is merely a coordinate
singularity. It can be removed by introducing ingoing Eddington-Finkelstein
coordinates $(\hat V, r, X, Y)$, defined by
\begin{equation}
\mathrm{d}\hat V = \mathrm{d}T + \frac{\mathrm{d}r}{f(r)}\,,
\end{equation}
in terms of which the metric becomes
\begin{equation}
\mathrm{d}s^2
 = -f(r)\,\mathrm{d}\hat V^{2}+ 2\,\mathrm{d}\hat V\,\mathrm{d}r+r^2\bigl(\mathrm{d}X^2 + \mathrm{d}Y^2\bigr)\,.
\end{equation}
The horizon is the null hypersurface at $r=r_h$, while the spacetime curvature
singularity lies at $r=0$.

We now introduce coordinates better adapted to our numerical scheme. In particular, we define
\begin{equation}
\hat V = \frac{\ell^2}{r_h} v,
\qquad
r = \frac{r_h}{z},
\qquad
X = \frac{\ell}{r_h} x,
\qquad\text{and}\qquad
Y = \frac{\ell}{r_h} y.
\end{equation}
In these coordinates, the line element becomes
\begin{equation}
\mathrm{d}s^2
 = \frac{\ell^2}{z^2}
   \left[
      -\left(1 - z^3\right)\mathrm{d}v^2
      - 2\,\mathrm{d}v\,\mathrm{d}z
      + \mathrm{d}x^2
      + \mathrm{d}y^2
   \right]\,,
   \label{eq:unibrane}
\end{equation}
with the horizon given by the null hypersurface $z=1$, the timelike boundary at $z=0$, and the spatial
coordinates obeying the identifications $(x,y)\sim (x+L_x,\,y+L_y)$, where
$L_x \equiv r_h L_X / \ell$ and similarly for $L_y$.

Next, we will generalize this setup by relaxing the translational symmetry in the $x$-direction and allowing the geometry to acquire explicit $v$-dependence. This will lead us to a more general ansatz suitable for describing time-dependent and spatially modulated configurations.

\subsection{Metric ansatz}
Following \cite{Balasubramanian_2014}, we generalise the line element (\ref{eq:unibrane}) to the ansatz
\begin{equation}
    \mathrm{d}s^2 = \frac{\ell^2}{z^2}\left\{ e^{2\beta}\left(-V\mathrm{d}v^2 - 2 \mathrm{d}v \mathrm{d}z\right) + e^{2\chi} \left[ \frac{1}{A^2}\left(\mathrm{d}x-U\mathrm{d}v\right)^2 + A^2\mathrm{d}y^2 \right] \right\}
\end{equation}
where the dynamical variables $\beta$, $V$, $\chi$, $U$, and $A$ depend only on $(v,z,x)$. This choice preserves the translational invariance in $y$, and effectively reduces this from a $3+1$ dimensional problem to $2+1$. This is important for the numerical method because a $2+1$ problem is computationally much cheaper, which allows for longer time evolutions. We also impose the symmetry condition $x\sim x+L$ where the parameter $L=L_x$ controls the horizon circumference. This modifies the spacetime topology to that of a toroidal AdS-Schwarzschild black hole, as studied in \cite{Dunn_2016, Dunn:2018let, Dunn:2018xdm}. We choose $L$ at the start of each time evolution, and taking the limit $L\to\infty$ would reproduce the planar case. This symmetry condition further simplifies the numerics as it gives the dynamical variables a natural decomposition as Fourier series.

The Einstein equation in 4D with a negative cosmological constant $\Lambda=-3/\ell^2$ is given by
\begin{equation}
    G_{ab} \equiv R_{ab}+\frac{3}{\ell^2}g_{ab}=0
\end{equation}

We require boundary conditions at the timelike boundary for this to be a well-posed initial value problem. We choose the standard Dirichlet boundary conditions with no sources. This corresponds to the conditions $\beta=\chi=U=0$ and $V=A=1$ at $z=0$.

We fix the remaining gauge freedom by choosing $\chi$ to take the form
\begin{equation}
    \chi(v,z,x)=\log{\left[ 1+z^3 \chi_3(v,x) \right]}
\end{equation}
and requiring the surface $z=1$ to be an apparent horizon. The apparent horizon condition corresponds to requiring the vanishing of the expansion of an outgoing null geodesic congruence, and in this case takes the simple form
\begin{equation}
    \left[\partial_v\chi+\frac{1}{2}V\left( \frac{1}{z}-\partial_z\chi \right)+\frac{1}{2}\,e^{-2\chi}\,\partial_x(e^{2\chi}U) \right]\bigg|_{z=1}=0\label{eq:AdS4_horizon_condition}
\end{equation}

By causality, information from behind the apparent horizon, \textit{i.e.} from $z>1$, cannot escape into $z\in[0,1]$. This allows us to excise this region and use $z\in[0,1]$ as our computational domain. Note that this gauge choice assumes not only the existence of an apparent horizon, but also that the apparent horizon is \textit{connected}. As \cite{Kehle:2022uvc} has strikingly demonstrated, connectedness cannot always be taken for granted. In our case, however, we study small perturbations of a single black‑brane spacetime that is manifestly non‑extremal, and therefore we do not expect this assumption to fail. In cases where a violation might occur, alternative gauges can be used - for example, simply requiring that $z=1$ be a null hypersurface which corresponds to fixing $V=0$ at $z=1$.

\subsection{Marching orders and new variables}\label{marching_orders}

The joy of using Bondi-Sachs coordinates is in the structure of the equations of motion and the constraint equations, which take the form of a nested set of first order ODEs on each hypersurface in the foliation. This nested structure, which is structurally the same as in \cite{Balasubramanian_2014, Crump:2025}, is as follows. Suppose that $A$ and $\chi$ are known on the hypersurface, then $G_{zz}=0$ can be solved for $\beta$. Next, we define the auxiliary variable $\Pi$ via
\begin{equation}
    \partial_zU = z^2 e^{2\beta-4\chi}A^2 \, \Pi\label{eq:AdS4_Pi_definition}
\end{equation}
We find $\Pi$ by solving $G_{zx}=0$, and then $U$ by solving \eqref{eq:AdS4_Pi_definition}. These require boundary conditions at $z=0$, for which we introduce the variable $U_3(v,x)$. The boundary conditions are then
\begin{equation}
    \lim_{z\to 0} \left[\frac{1}{z^3}U(v,z,x)\right] = \frac{1}{3}\,\Pi(v,z=0,x) = U_3(v,x)
\end{equation}
We introduce two more auxiliary variables, $d_t\chi$ and $d_tA$, defined via
\begin{align}
    \partial_v\chi &= z^2 e^{-2\chi} d_t\chi + \frac{1}{2}\left(V-1\right)\left(\partial_z\chi-\frac{1}{z}\right)\label{eq:AdS4_dtchi_definition} \\
    \partial_vA &= z e^{-\chi}A \,d_tA + \frac{1}{2} V \partial_zA\label{eq:AdS4_dtA_definition}
\end{align}
We can use $g^{ab}G_{ab}=0$, $(a,b)\in(x,y)$, to find $d_t\chi$ and then $G_{xx}=0$ to find $d_tA$. This requires a boundary condition for $d_t\chi$. One option is to use the apparent horizon condition at $z=1$, but instead we choose to introduce the variable $V_3(v,x)$ to define the boundary condition at $z=0$ given by
\begin{equation}
    d_t\chi(v,z=0,x) = -\frac{1}{2} + 2 \chi_3(v,x)+\frac{1}{2} V_3(v,x)
\end{equation}

By defining the boundary variables $U_3$ and $V_3$ in this way, they have physical significance. Via the AdS/CFT correspondence, the bulk gravitational dynamics with a toroidal black hole horizon in AdS$_4$ corresponds to the dynamics of a boundary QFT on $\mathbb{R}\times\mathbb{T}^2$ at finite temperature. In the long wavelength limit, the boundary dynamics becomes hydrodynamics (\textit{i.e.} the fluid/gravity correspondence). The variables $U_3$ and $V_3$ correspond to the expectation values $\langle T_{vx}\rangle$ and $\langle T_{vv}\rangle$ of the QFT stress tensor $T_{\mu\nu}$, up to multiplicative factors. 

From a practical perspective, the prefactors that multiply $\Pi$, $d_t\chi$, and $d_tA$ in the definitions \eqref{eq:AdS4_Pi_definition}, \eqref{eq:AdS4_dtchi_definition}, and \eqref{eq:AdS4_dtA_definition} are chosen so that the differential operators in their respective ODEs are independent of $x$. This simplifies the numerics because it means that the same discretised operator can be used at each $x$.

The final differential equation to be solved on the hypersurface is a second order elliptic equation at $z=1$ for the horizon value of $V$. This comes from the combination $G^{z}_{\,\,v}+U G^{z}_{\,\,x}=0$, where $G_{zz}=0$ and the apparent horizon condition \eqref{eq:AdS4_horizon_condition} can be used to eliminate terms proportional to $V^2$, $\partial_zV$, and $\partial_vV$. The boundary conditions for this equation come from the periodicity condition $x\sim x+L$, and are automatically satisfied when using a Fourier spectral method. Equipped with the horizon value of $V$, the apparent horizon condition \eqref{eq:AdS4_horizon_condition} can be rearranged to give $\partial_v \chi_3$. Our gauge choice $\chi=\log{(1+z^3\chi_3)}$ means that we know $\partial_v\chi$ everywhere now that we know $\partial_v\chi_3$. Knowing $\partial_v\chi$, we know $V$ everywhere by rearranging the definition of $d_t\chi$. Finally, we can find the time derivative $\partial_vA$ by rearranging the definition of $d_tA$. 

The only remaining time derivatives to be found are $\partial_vU_3$ and $\partial_vV_3$, which can be found by performing a boundary expansion in $z$ of the Einstein equation. In particular, this requires expanding the unused $G_{vv}$ and $G_{vx}$ equations. We find
\begin{align}
    \partial_vU_3 &= \frac{1}{3}\partial_xV_3 + 2 \partial_x A_3 \\
    \partial_vV_3 &= \frac{3}{2} \partial_x U_3
\end{align}
\noindent where $A_3(v,x)$ is the first subleading term in the boundary expansion of $A$, such that $A(v,z,x)=1+z^3A_3(v,x)+\mathcal{O}(z^4)$.

Initial data is specified by choosing $A$, $U_3$, $V_3$, and $\chi_3$ on the initial hypersurface $v=0$, and then the marching orders described produce the time derivatives of these four variables. This allows us to implement a numerical time evolution scheme (we choose RK4) to find these variables on the next hypersurface in the discretised foliation. In practice, we only specify $A$, $U_3$, and $V_3$ as initial data and then find $\chi_3$ numerically by requiring that $z=1$ be an apparent horizon.

We define new variables $Q_1$ and $p_i$, $1\leq i \leq 6$, for use in the numerics in order to algebraically remove divergences as $z \to 0$ that appear in terms such as $(A-1)/z^3$ and $U/z^3$. These new variables are
\begin{subequations}
\begin{align}
    A(v,z,x) &= 1 + Q_1(v,z,x) z^3 \\
    \beta(v,z,x) &= -\chi_3(v,x) z^3 + \left[ p_1(v,z,x) - \frac{3}{4} A_3(v,x)^2 - \chi_3(v,x)^2\right] z^6 \\
    \Pi(v,z,x) &= p_2(v,z,x) + 3 U_3(v,x)\\
    U(v,z,x) &= \Big[ p_3(v,z,x) + U_3(v,x) \Big]z^3  \\
    d_t\chi(v,z,x) &= p_4(v,z,x) -\frac{1}{2} +2 \chi_3(v,x) + \frac{1}{2} V_3(v,x) \\
    d_tA(v,z,x) & = p_5(v,z,x) - \frac{3}{2} A_3(v,x) \\
    V(v,z,x) &= (1-z^3) + \Big[ 4 \chi_3(v,x) + V_3(v,x) \Big] z^3 + p_6(v,z,x) z^4
\end{align}
\end{subequations}
This also simplifies the boundary conditions for the ODEs, which are now just $p_i(v,z=0,x)=0$ for $1\leq i\leq 5$.

For our numerical setup, we use a Fourier spectral method in the $x$-direction with equispaced collocation points. In the $z$-direction, we use a discontinuous Galerkin (DG) method with Gauss-Legendre-Lobatto collocation points in each element. The DG method uses an upwind flux for the hypersurface ODEs and a Lax-Friedrichs flux for the time evolution equations.

\subsection{Prediction of late time behaviour}
\begin{figure}[t]
    \centering
    \includegraphics[width=0.9\linewidth]{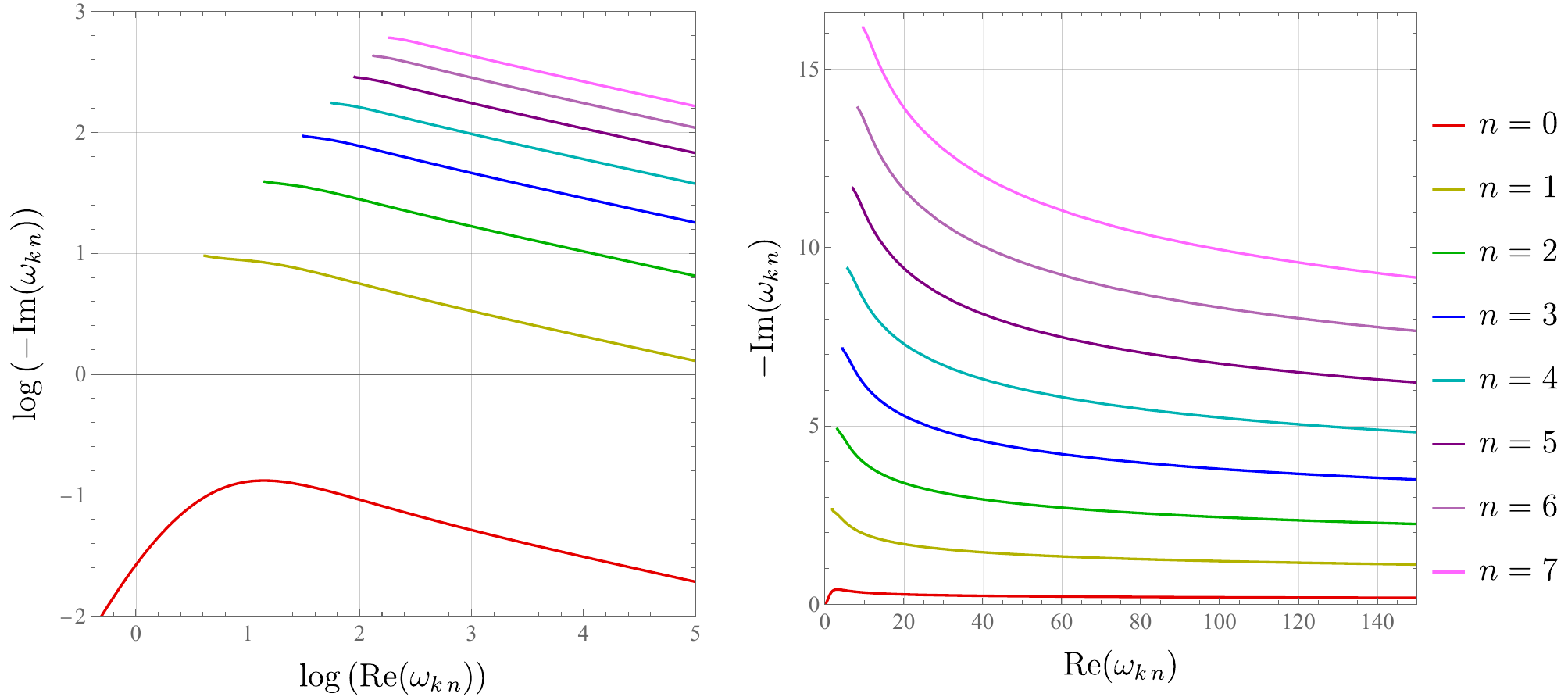}
    \caption{\textit{(Left)} Plot of the log of the decay rate $-\mathrm{Im}(\omega_{k\,n})$ against the log of the frequency $\mathrm{Re}(\omega_{k\,n})$ up to overtone $n=7$ for QNMs of the planar black hole in AdS$_4$. For each overtone, modes are continuous lines in contrast to the discrete spectra of black holes with spherical horizon topology. Only the $n=0$ mode line connects to $\omega_{k\,n}=0$ at $k=0$, and the $n>0$ modes are gapped. \textit{(Right)} The plot without taking the log of each axis, shown to highlight the spacing of the mode lines.}
    \label{fig:AdS4_QNM_frequencies_and_decay_rates}
\end{figure}

\begin{figure}[t]
\centering
    \includegraphics[width=0.9\linewidth]{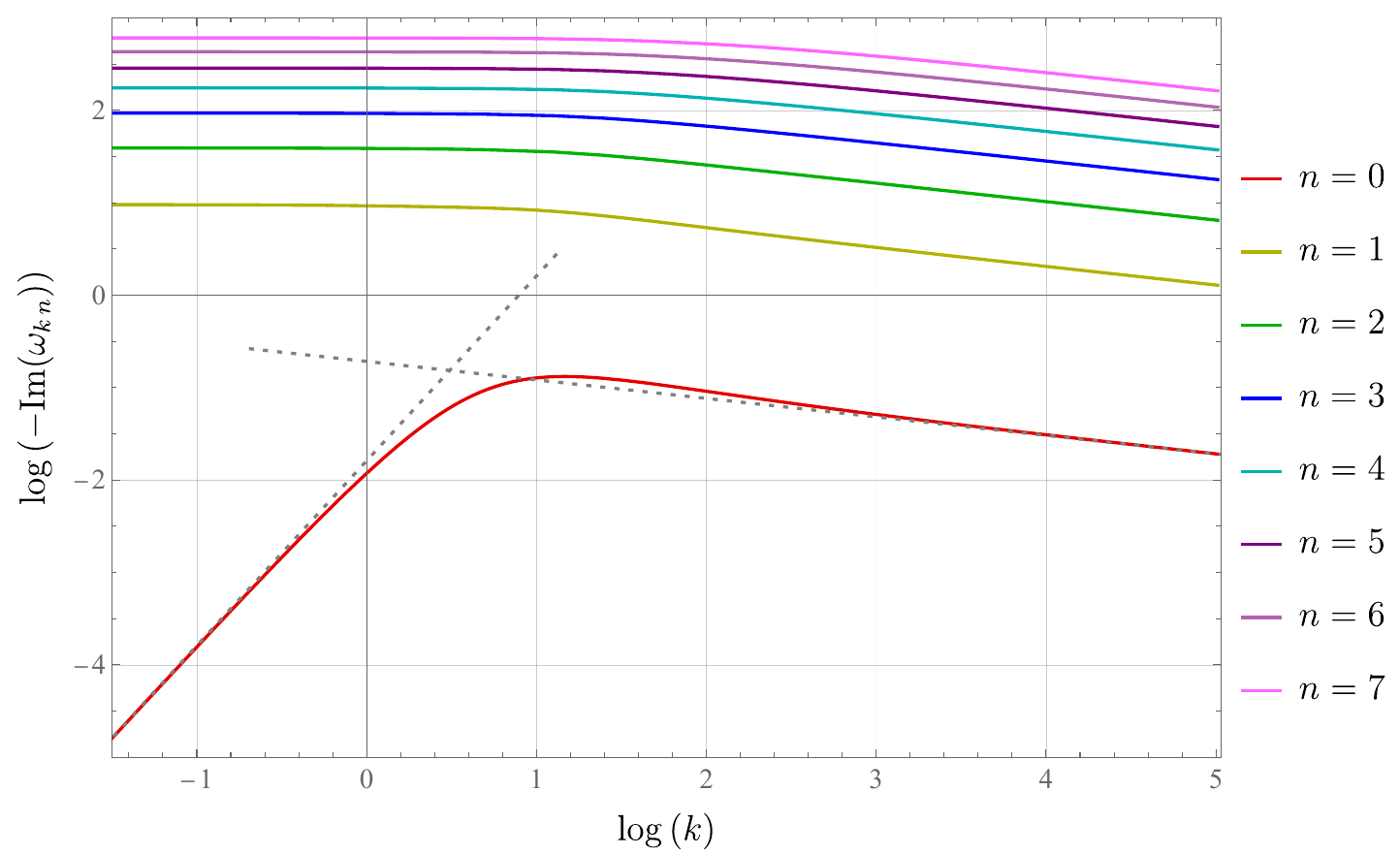}
    \caption{Log-log plot of the QNM decay rate $-\mathrm{Im}(\omega_{k\,n})$ against wavenumber $k$, for mode lines up to overtone $n=7$. All mode lines become linear asymptotically as $k\to \infty$ which indicates power law scaling, each scaling with $k^{-1/5}$. As $k\to 0$ the $n=0$ mode scales with $k^2$, which represents the diffusive hydrodynamics regime at the boundary. The two asymptotes for the $n=0$ mode line are shown as dashed grey lines.}
    \label{fig:AdS4_QNM_log_k_log_omega}
\end{figure}

To predict the late time behaviour of perturbations, we first take a closer look at the QNM spectrum. QNMs of the black brane are finite energy solutions to the linearised Einstein equation that characterise the time and spatial dependence of small perturbations. At late times we expect the initial perturbation, which may be large, to gradually settle down to the equilibrium state (the stationary planar black hole). Therefore we expect that at late times the solution can be well-approximated by a superposition of QNMs. The QNM ansatz has $v$ and $x$ dependence of the form $e^{-{\rm i}\,\omega_{k\,n}\,v +{\rm i}\,k\,x}$ and we find solutions that are labelled by an overtone number $n=0,1,2,...$ and the wavenumber $k$. $\omega_{k\,n}$ is complex, with the real part corresponding to the frequency of the QNM and the imaginary part corresponding to the decay rate. Without imposing the periodicity condition $x\sim x+L$, the modes at each $n$ are continuous lines. Fig.~\ref{fig:AdS4_QNM_frequencies_and_decay_rates} shows the decay rates $-\mathrm{Im}(\omega_{k\,n})$ of these mode lines up to $n=7$ plotted against the frequencies $\mathrm{Re}(\omega_{k\,n})$, and Fig.~\ref{fig:AdS4_QNM_log_k_log_omega} shows a log-log plot of the decay rates $-\mathrm{Im}(\omega_{k\,n})$ plotted against the wavenumber $k$.


We make several observations from Fig.~\ref{fig:AdS4_QNM_log_k_log_omega}. The first observation is that the lowest overtone $n=0$ has a much slower decay rate than the higher overtones at each wavenumber $k$. This means that we will only consider the $n=0$ contribution to the late time behaviour. The second observation is that all of the mode lines become linear asymptotically as $k\to \infty$ on the log-log plot, indicative of a power law of the form $-\mathrm{Im}(\omega_{k\,n})\sim \gamma_n\,k^{-\sigma}$ for each overtone (each with different $\gamma_n>0$ but the same $\sigma>0$). Indeed, the analysis in \cite{Festuccia:2008zx} predicts an asymptotic scaling of this form, where $\sigma$ is found to be $\frac{d-2}{d+2}$ in AdS$_{d+1}$. Explicitly, in the case of the AdS$_4$ black brane this corresponds to $\sigma=\frac{1}{5}$ which is in empirical agreement with the slopes of the asymptotes in Fig.~\ref{fig:AdS4_QNM_log_k_log_omega}. The third observation is that $n=0$ is the only mode line that connects to $\omega_{k\,n}=0$ at $k=0$, and for small $k$ has a decay rate that scales with $k^2$. In the language of the boundary theory, this scaling represents the well-studied long wavelength low frequency hydrodynamic shear mode, characterised by diffusive behaviour. The last observation is that for $n=0$ there is a $k=k_{\mathrm{max}}$ with a maximum decay rate (from the plot we see that this is around $\log{(k_{\mathrm{max}})}\approx 1$). For $k<k_{\mathrm{max}}$ the decay rate increases monotonically from zero, and for $k>k_{\mathrm{max}}$ the decay rate decreases monotonically towards zero. This represents the change from long wavelength hydrodynamic behaviour to the geometric optics regime where high momentum modes become increasingly localised near the AdS boundary.

\begin{figure}[t]
\centering
    \includegraphics[width=0.9\linewidth]{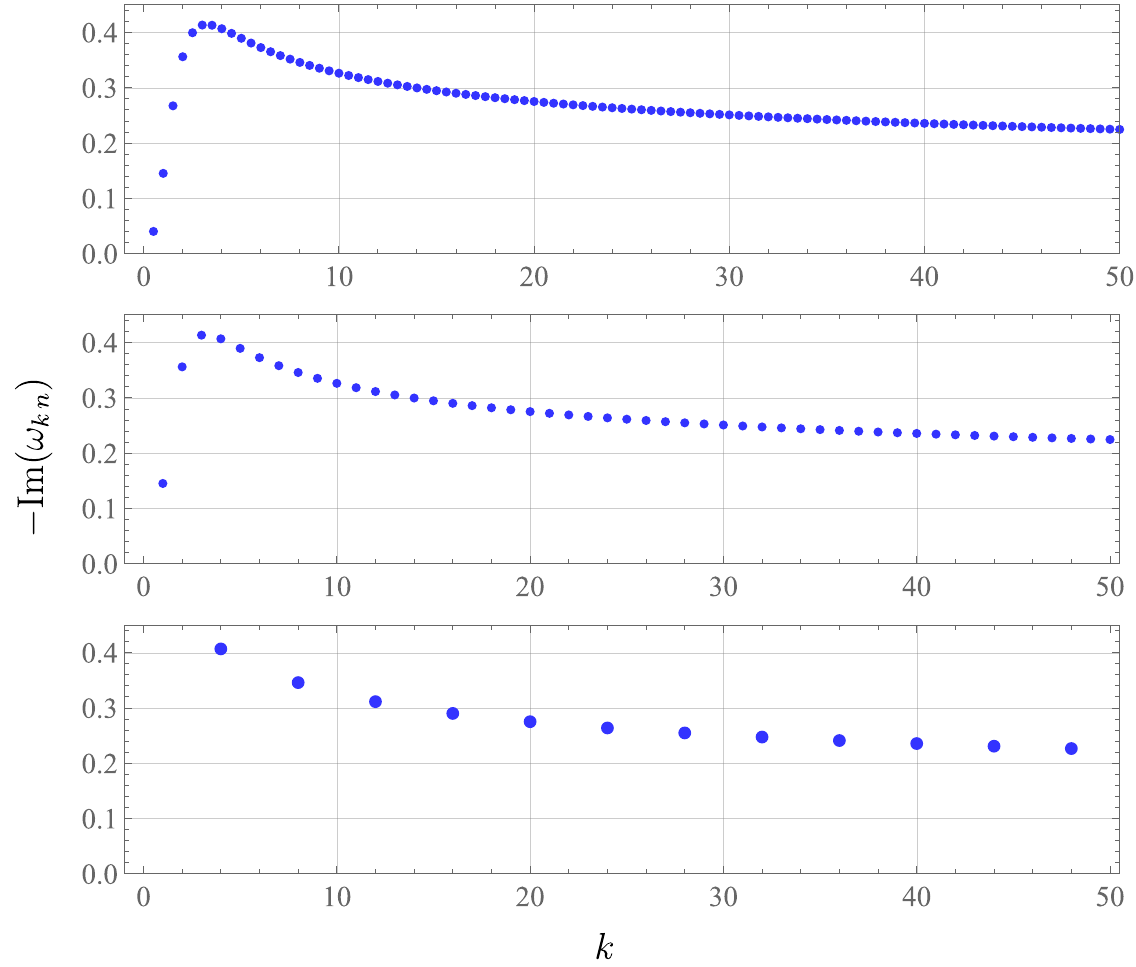}
    \caption{Plot of the decay rate against wavenumber of the $n=0$ mode line after imposing the periodicity $x\sim x+L$. The spectrum becomes discrete. The top plot shows $L=4\pi$, the middle plot shows $L=2\pi$, and the bottom plot shows $L=\pi/2$.}
    \label{fig:AdS4_L_comparison_decay_rates}
\end{figure}

Imposing the periodicity condition $x\sim x+L$ changes the picture. Now only a discrete set of modes survive, restricting the mode lines to their values at $k=2\pi m/L$, $m\in \mathbb{Z}$, with the first non-zero mode at $k=2\pi/L$. This splits the behaviour of the time evolution into two classes depending on $L$. If $L<2\pi/k_{\mathrm{max}}$, then from the last observation we see that the decay rates of the modes with non-zero $k$ strictly decrease with $k$. This allows us to predict the late time behaviour by approximating the mode decay rates by the asymptotic scaling $-\mathrm{Im}(\omega_{k\,n}) =  \gamma_n\,k^{-\sigma}$, and the true behaviour should gradually approach the prediction. In the other case, where $L>2\pi/k_{\mathrm{max}}$, there will be two regimes. There will be a finite set of long lived modes with low wavenumbers $2\pi/L\leq k <k_{\mathrm{max}}$, with $k=2\pi/L$ having the slowest decay rate. These will dominate the time evolution until they have decayed enough to be smaller than the modes at very large $k\gg k_{\mathrm{max}}$ with decay rates even lower than theirs. Beyond that point, the behaviour can be predicted by considering a tail of modes with decay rates following $-\mathrm{Im}(\omega_{k\,n}) =  \gamma_n\,k^{-\sigma}$. These two classes of behaviour are illustrated in Fig.~\ref{fig:AdS4_L_comparison_decay_rates}, which shows the decay rates for $L=2\pi$ and $L=4\pi$, both of which have $L>2\pi/k_{\mathrm{max}}$, and for $L=\pi/2$ which has $L<2\pi/k_{\mathrm{max}}$.

We now make a quantitative prediction for the late time behaviour of both classes, in the regime where the behaviour is dominated by the tail of the spectrum at large $k$. We track the norm of $V_3(v,x)$ because $V_3$ has a physical interpretation as the energy density of the boundary theory. This will act as a surrogate for predictions about other variables, for example the time dependence of curvature invariants integrated over spatial slices. We can decompose $V_3$ into Fourier modes as
\begin{equation}
    V_3(v,x)=\sum_{m=-\infty}^{\infty} \tilde{V}_{3 \, m}(v) \exp{\left(\frac{2\pi m x}{L}{\rm i}\right)}
\end{equation}\label{eq:Fourier_decomposition_of_V3}
\noindent where $\tilde{V}_{3 \, -m} = \tilde{V}_{3 \, m}^{*}$ as $V_3$ is real. Then we define the norm $\| V_3\|^2(v)$ as
\begin{equation}
    \| V_3\|^2(v) = \frac{1}{L}\int_{0}^{L} |V_3(v,x)|^2 \mathrm{d}x = |\tilde{V}_{3 \, 0}(v)|^2 + 2\sum_{m=1}^{\infty} |\tilde{V}_{3 \, m}(v)|^2 \label{eq:V3_norm_definition}
\end{equation}
where the second equality follows from Parseval's theorem. The mode $\tilde{V}_{3 \, 0}$ is a conserved quantity, corresponding to the conservation of energy of the boundary theory. This means that choosing initial data with $\tilde{V}_{3 \, 0}=0$ will guarantee that it remains zero throughout the time evolution. In the solutions discussed in the results, we make the stronger choice $V_3=0$ on the initial hypersurface so that the modes $m\geq1$ are switched on through early time nonlinear interactions.

Next, we make an assumption about the spectrum of $V_3$. We assume that at some time $v_0$ the mode amplitudes scale exponentially with $m$ and can be approximated by
\begin{equation}
    |\tilde{V}_{3 \, m}(v_0)| \approx \exp{\left( - a m+b \right)}
\end{equation}
\noindent for some real $a,b$ with $a>0$. This is a reasonable approximation for a solution with real-analytic initial data as analyticity implies $|\tilde{V}_{3 \, m}(v_0)| \leq \exp{\left( - a m+b \right)}$ for some $a,b$, assuming that the time evolution has preserved the smallness of the analytic norm.

We make the further assumption that beyond some late time $v\gg v_0$ the solution has decayed to become small enough that the mode behaviour can be well-approximated by QNMs. This approximates the time dependence as
\begin{equation}
    \tilde{V}_{3 \, m}(v) \approx \tilde{V}_{3 \, m}(v_0) \exp{\left[ -{\rm i}\,\omega_m (v-v_0) +{\rm i}\,\phi_m\right]}
\end{equation}
\noindent for phases $\phi_m$ and where $\omega_m\equiv \omega_{k\,0}$ are the complex QNM frequencies for the $n=0$ branch discussed previously. The final approximation we make is that the decay rates scale as $-\mathrm{Im}(\omega_m) =  \rho  m^{-\sigma}$ where $\sigma=\frac{1}{5}$ and $\rho>0$. This approximation rests on the assumption that we are in the late time regime where the spectrum is dominated by the tail $m\gg 1$.

Putting this all together and dropping the phase and $v_0$ dependencies to focus on the character of the time dependence, we find
\begin{equation}
    \| V_3\|^2(v) \approx  \sum_{m=1}^{\infty} \exp{\left(-2am+2b-2\rho m^{-\sigma} v \right)}\label{eq:V3_norm_approximation_1}
\end{equation}
where the prefactor of 2 in Eq.~\eqref{eq:V3_norm_definition} has been absorbed into $b$. We now explore how this sum scales with time. The most crude extraction of the time dependence comes from
\begin{equation}
    \log{\left( \| V_3\|^2(v) \right)} \approx \max_{m} \left( -2am+2b-2\rho m^{-\sigma} v \right)
\end{equation}
which yields the scaling
\begin{equation}
    \| V_3\|^2(v) \approx \exp{\left[2b -2\rho (\sigma+1)\left( \frac{a}{\rho \sigma} \right)^{\frac{\sigma}{\sigma+1}} \, v^{\frac{1}{\sigma+1}} \right]}
\end{equation}
To improve upon this, we note that \eqref{eq:V3_norm_approximation_1} can be written as
\begin{align}
    \| V_3\|^2(v) &\approx  e^{2b} \sum_{m=1}^{\infty} \exp{\left\{v^{\frac{1}{\sigma+1}}\left[-2a\left(v^{-\frac{1}{\sigma+1}}m\right)-2\rho \left(v^{-\frac{1}{\sigma+1}}m\right)^{-\sigma} \right]\right\}}\label{eq:factoring_the_sum}\\
    &\approx e^{2b} \int_{1}^{\infty} \exp{\left[v^{\frac{1}{\sigma+1}}\left(-2au-2\rho u^{-\sigma} \right)\right]} v^{\frac{1}{\sigma+1}} \mathrm{d}u
\end{align}
As $v$ becomes large at late times, this integral is perfectly poised for Laplace's method \cite{Laplace_method}. This results in
\begin{equation}
    \| V_3\|^2(v) \approx e^{2b} \sqrt{\frac{\pi}{\rho \sigma(\sigma+1) ( \frac{a}{\rho \sigma})^{\frac{\sigma+2}{\sigma+1}}}} \, \,v^{\frac{1}{2(\sigma+1)}} \exp{\left[ -2\rho (\sigma+1)\left( \frac{a}{\rho \sigma} \right)^{\frac{\sigma}{\sigma+1}} \, v^{\frac{1}{\sigma+1}} \right]}
\end{equation}
To highlight the form of the predicted time dependence, we write this in the form
\begin{equation}
    \log{\left(\| V_3\|^2(v)\right)} \approx -c_1 \exp{\left(\frac{1}{\sigma+1} \log{v}\right)}+\frac{1}{2(\sigma+1)} \log{v}+c_2
\end{equation}
\noindent where $c_1>0$, $c_2\in\mathbb{R}$. For concreteness, with $\sigma=\frac{1}{5}$ this is the subexponential decay given by
\begin{equation}
    \log{\left(\| V_3\|^2(v)\right)} \approx {\displaystyle -c_1 v^{\frac{5}{6}}}+\textstyle\frac{5}{12} \log{v}+c_2 \label{eq:subexponential_late_time_prediction}
\end{equation}

We can also make predictions about the evolution of the spatial profile of the perturbation. So far, we have approximated the Fourier spectrum of $V_3$ as
\begin{equation}
    |\tilde{V}_{3 \, m}(v)| \approx \exp{\left( -am+b-\rho m^{-\sigma} v \right)}\label{eq:V3_spectrum_coefficients}
\end{equation}
which has its peak at $m_{\mathrm{max}}(v)=(\rho \sigma v/a)^{\frac{1}{\sigma+1}}$. Expanding the exponent of \eqref{eq:V3_spectrum_coefficients} around this peak to second order (as in Laplace's method previously), we find
\begin{multline}
    |\tilde{V}_{3 \, m}(v)| \approx \exp{\left\{ b - v^{\frac{1}{\sigma+1}}  \left[ \rho(\sigma+1)   \left( \frac{a}{\rho \sigma} \right)^{\frac{\sigma}{\sigma+1}}\right] \right\} }\\
    \times\exp{\left\{   -\frac{1}{2 v^{\frac{1}{\sigma+1}}} (m-m_{\mathrm{max}}(v))^2 \left[ \rho \sigma (\sigma+1) \left( \frac{a}{\rho \sigma} \right)^{\frac{\sigma+2}{\sigma+1}} \right] \right\}}
\end{multline}
The key observation we make here is that the spectral width of this Gaussian wavepacket approximation \textit{increases} with time, scaling with $v^{\frac{1}{2(\sigma+1)}}$. In physical space, this corresponds to the characteristic length scale of the perturbation \textit{decreasing} with time, scaling as $v^{-\frac{1}{2(\sigma+1)}}$. This focusing behaviour is in stark contrast to the diffusion seen in the hydrodynamics regime.

\section{Results}

\subsection{A small black hole: $L=2\pi/3$}

We begin by studying a black hole with a small horizon length scale $L=2\pi/3$. This is small in the sense that $L<2\pi/k_{\mathrm{max}}$ so the QNM decay rates strictly decrease with wavenumber $k$, leaving no trace of the hydrodynamic scaling near $k=0$. Whilst an even smaller black hole would have a QNM spectrum that better exhibits the asymptotic power law scaling as $k\to\infty$, it would also decay more slowly and oscillate with higher frequencies - requiring longer evolutions and higher resolutions. The choice $L=2\pi/3$ strikes a balance between prediction and computational feasibility.

For initial data on $v=0$, we choose real-analytic profiles given by
\begin{subequations}
\begin{align}
    Q_1(0,z,x) &=  0.01 \left[ \frac{\cos{(3x)}}{1.1+\cos{(3x)}} + \sin^3{(3x)} \right]\\
    V_3(0,x) &=0\\
    U_3(0,x) &= 0
\end{align}
\end{subequations}
The choice of $\chi_3(0,x)$ is fixed by requiring an apparent horizon at $z=1$. The choice of initial data is essentially arbitrary other than the requirement of analyticity and periodicity with period $L$. The reason for this choice is that it leads to $x$-dependence with a shallow slope of the Fourier mode amplitudes against $k$ on a logarithmic scale. This helps to keep a large number of modes numerically resolvable during the time evolution. The choice $V_3(0,x)=0$ ensures that the conserved $k=0$ mode is zero throughout the evolution. 

For the numerics, we use a Fourier spectral method in the $x$-direction with $128$ equispaced collocation points. In the $z$-direction, we use a discontinuous Galerkin method with $10$ equally sized elements, each with 11 Gauss-Legendre-Lobatto collocation points. RK4 is used as the time evolution scheme. We use a timestep $\Delta v=0.0003$ (which we observe to be close to the threshold CFL condition for stability) and evolve up to $v=146$. Beyond this time, the perturbation becomes too small to accurately resolve with machine precision computation.

\begin{figure}[H]
\centering
\begin{tabular}{cc}
     \includegraphics[width=0.85\linewidth, align=c]{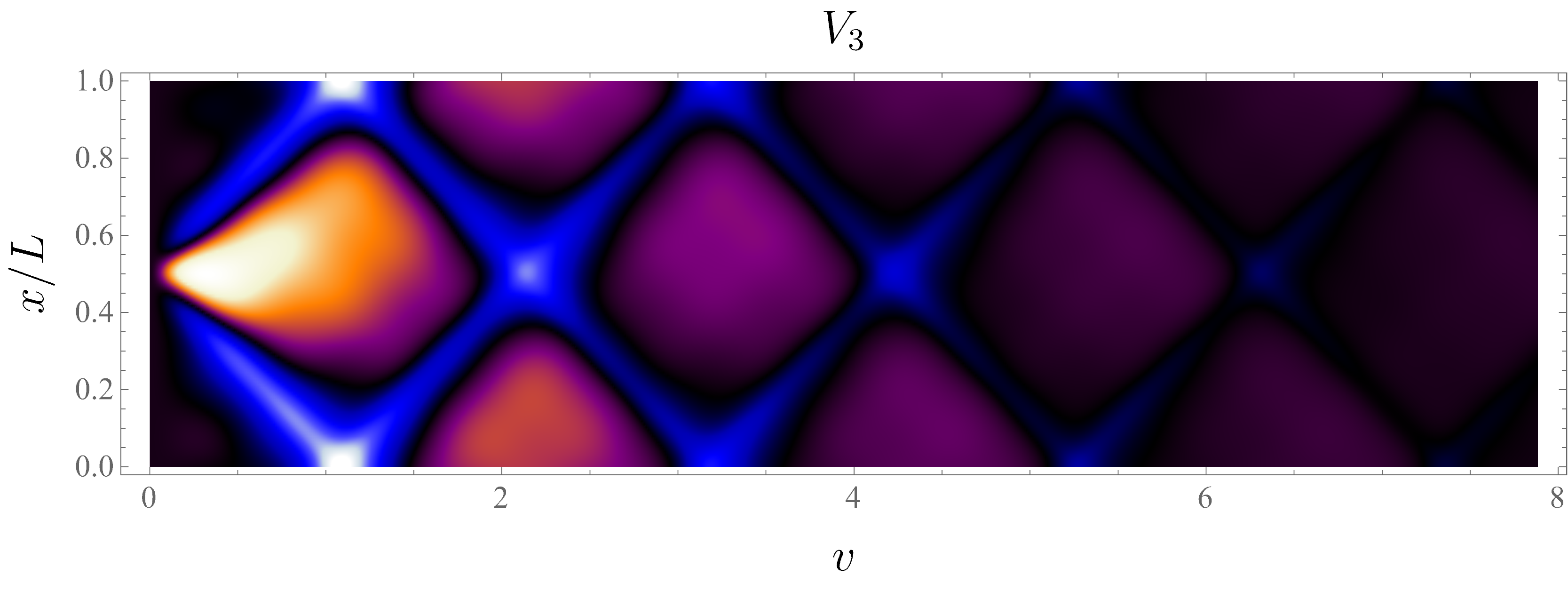} & \includegraphics[width=0.055\linewidth, align=c]{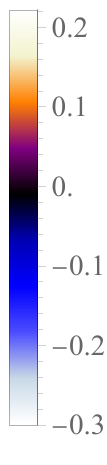}  \\
     \includegraphics[width=0.85\linewidth, align=c]{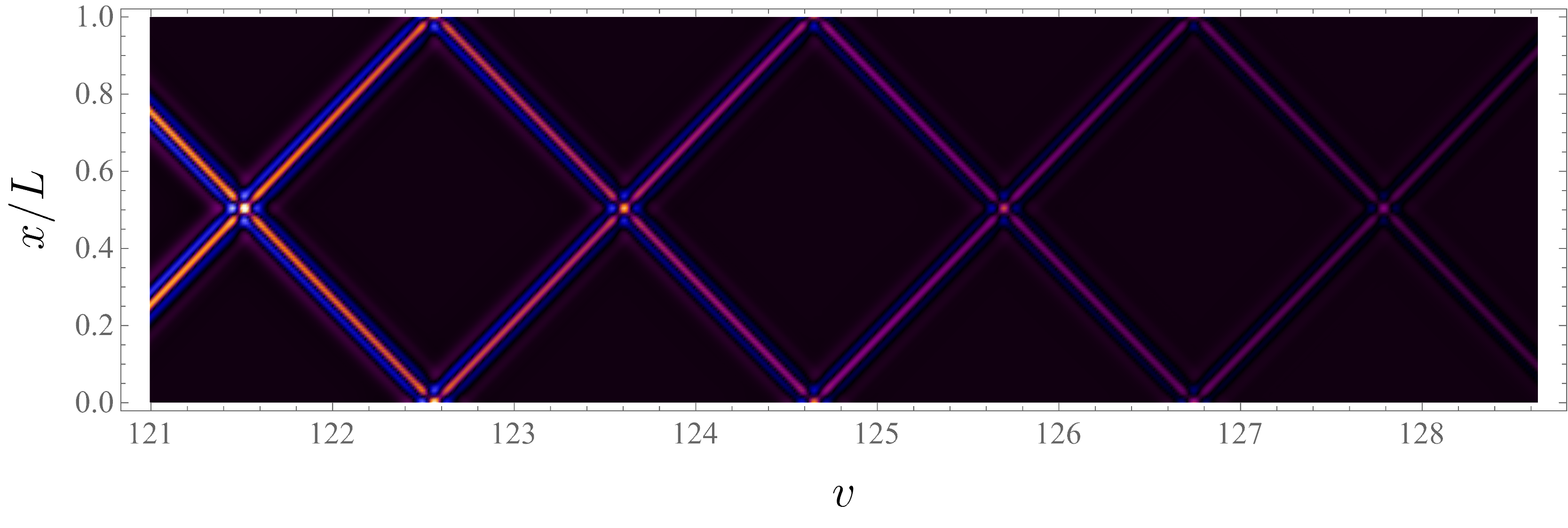} & \includegraphics[width=0.08\linewidth, align=c]{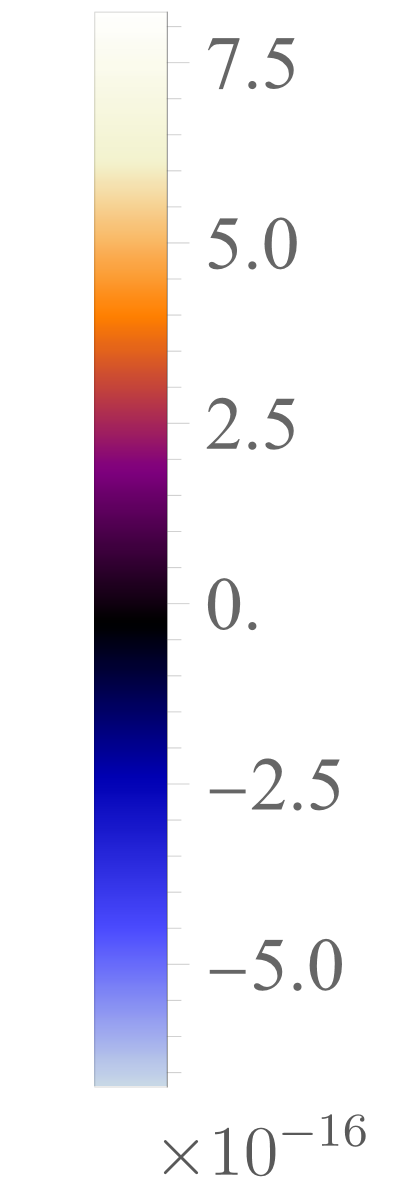} \\[17ex]
     \includegraphics[width=0.85\linewidth, align=c]{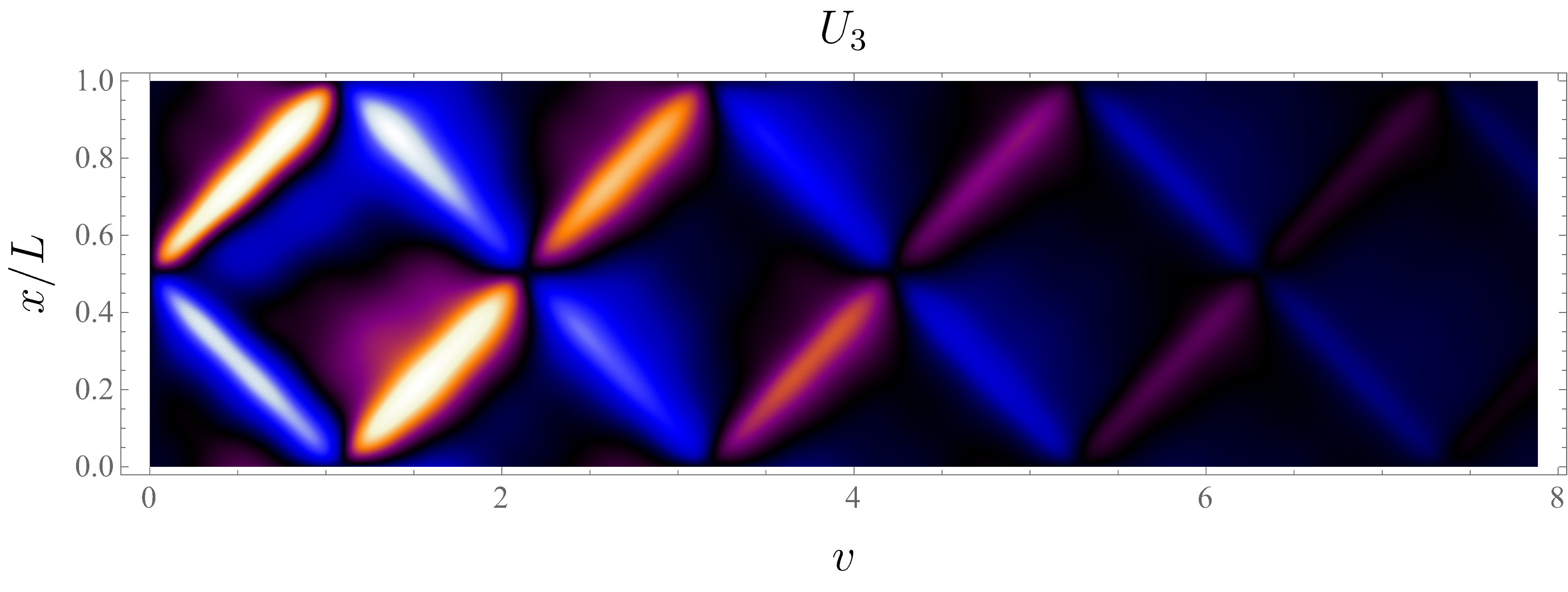} & \includegraphics[width=0.06\linewidth, align=c]{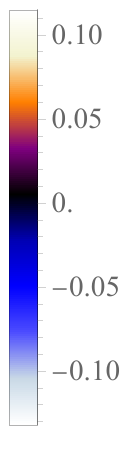}  \\
     \includegraphics[width=0.85\linewidth, align=c]{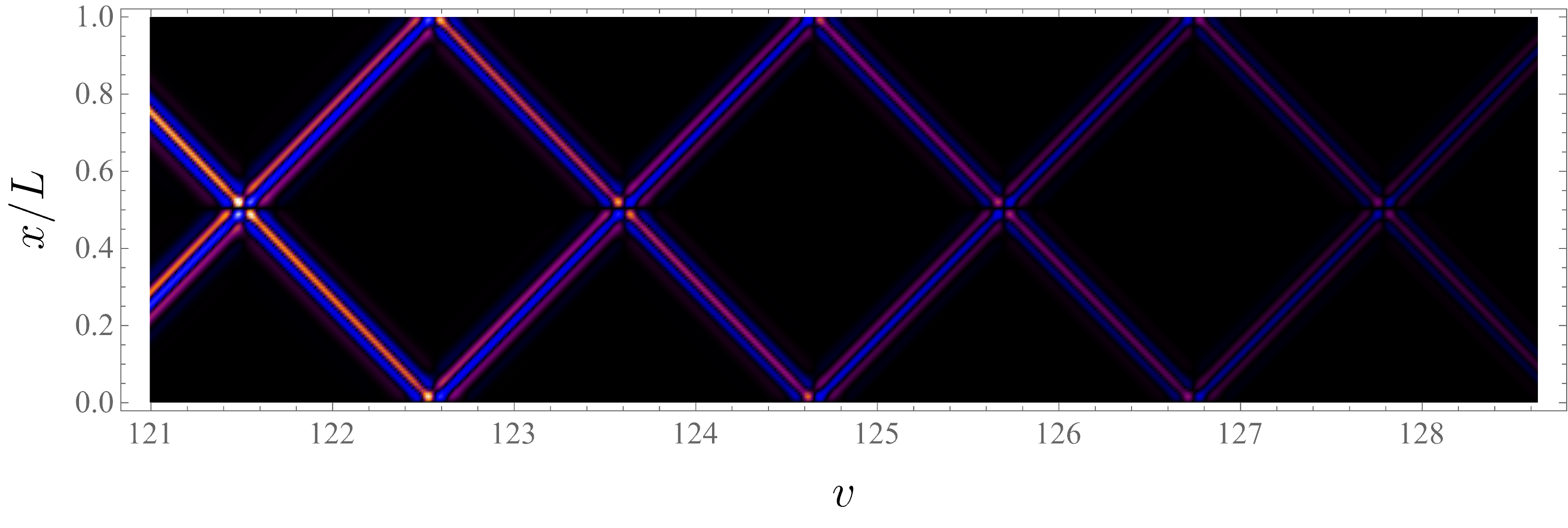} & \includegraphics[width=0.085\linewidth, align=c]{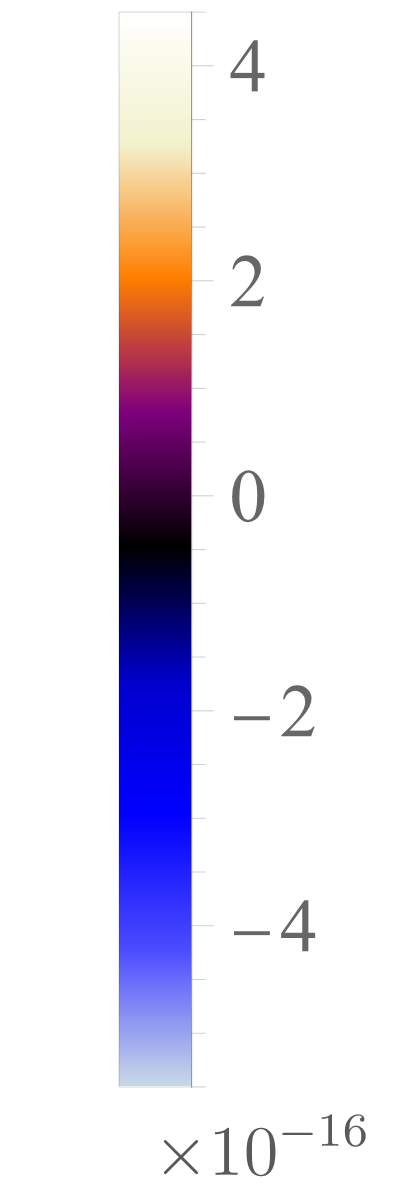} 
\end{tabular}
    \caption{\textit{(Top)} Time evolution of the boundary energy density $V_3$ at early times and at late times. \textit{(Bottom)} Time evolution of the boundary momentum density $U_3$ during the same early and late time intervals. Both $V_3$ and $U_3$ become increasingly localised as time progresses.}
    \label{fig:V3_and_U3_small_black_hole}
\end{figure}

In \autoref{fig:V3_and_U3_small_black_hole}, we show the early and late time evolution of the boundary energy density $V_3$ and momentum density $U_3$. Firstly, we make some qualitative observations. As expected, the solution does significantly decay - from an initial scale of $\sim10^{-1}$ to $\sim10^{-16}$ by the end of the runtime. We also observe that the solution appears to become more and more spatially localised during the evolution. This is what we expect, and this narrowing of the wavepacket is due to the broadening of the Fourier spectrum of the perturbation with time. This also highlights that the predicted late time behaviour, due to the spectral tail at large $k$, corresponds to the geometric optics regime of the perturbation - which aligns with the observed short wavelength behaviour.

\begin{figure}[H]
\centering
\begin{tabular}{ccc}
     \includegraphics[width=0.27\linewidth, align=c]{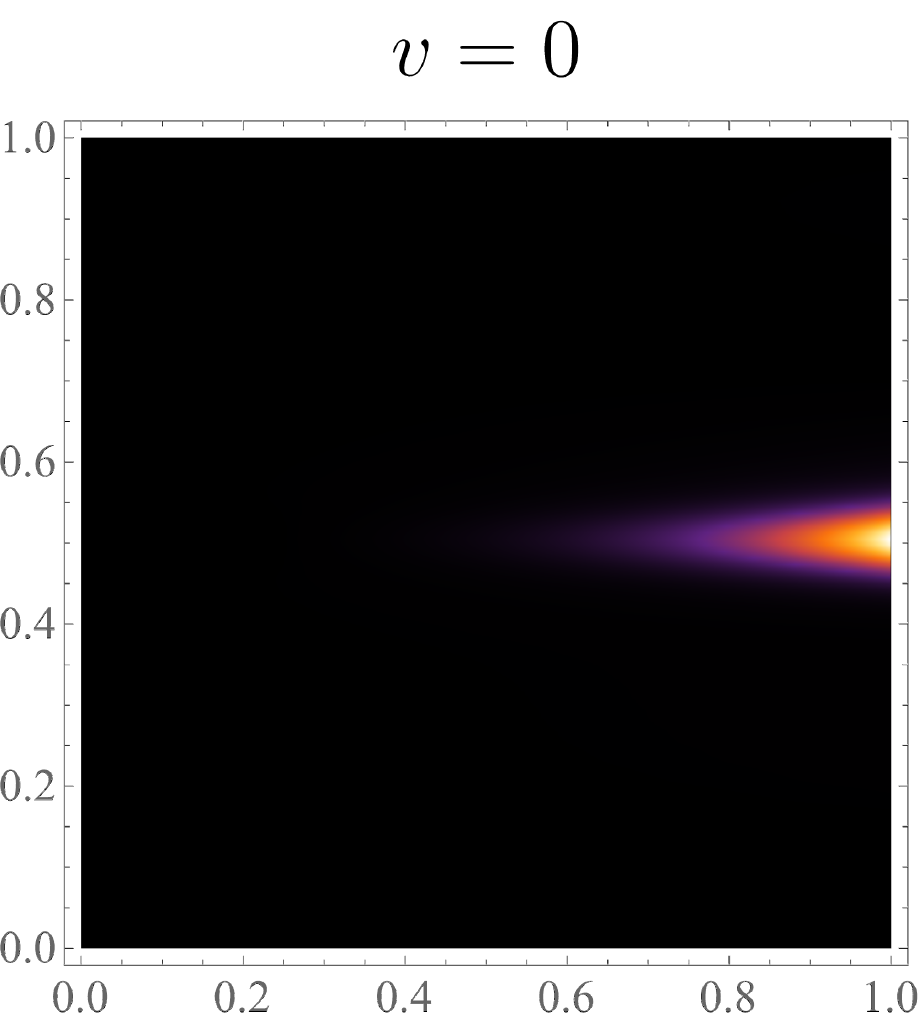} &
     \includegraphics[width=0.27\linewidth, align=c]{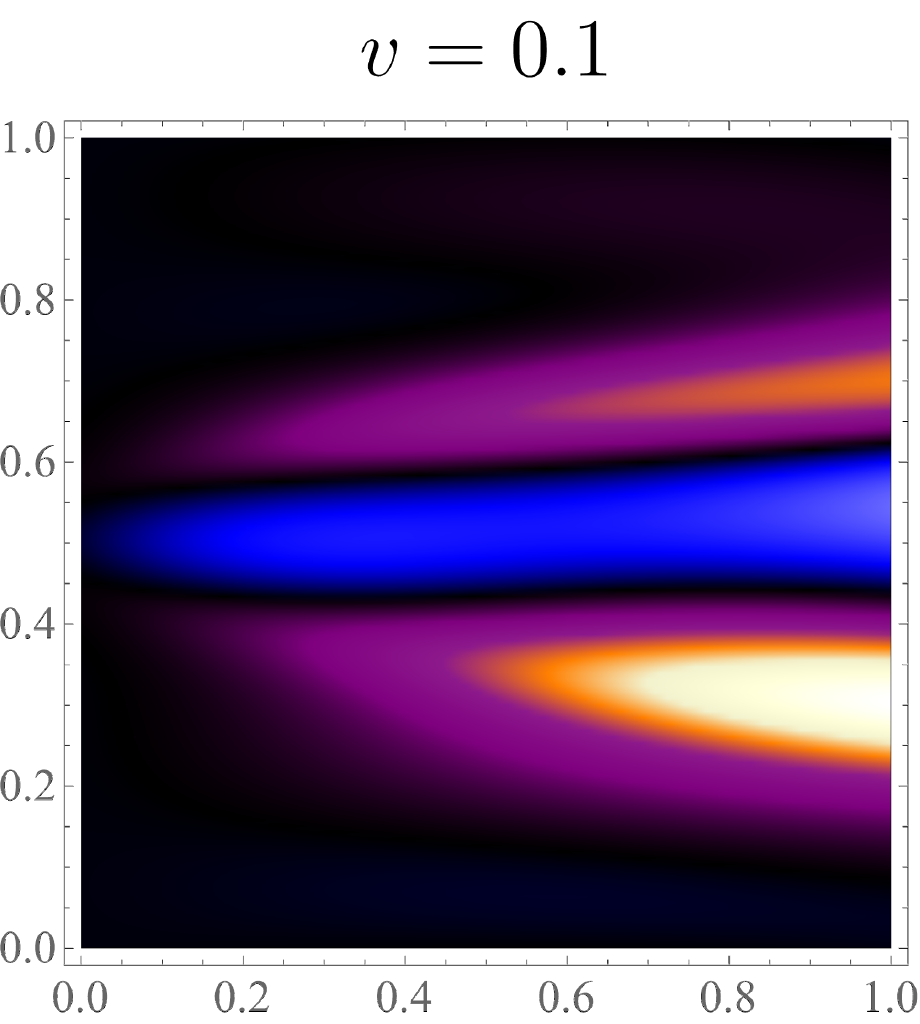} &
     \includegraphics[width=0.27\linewidth, align=c]{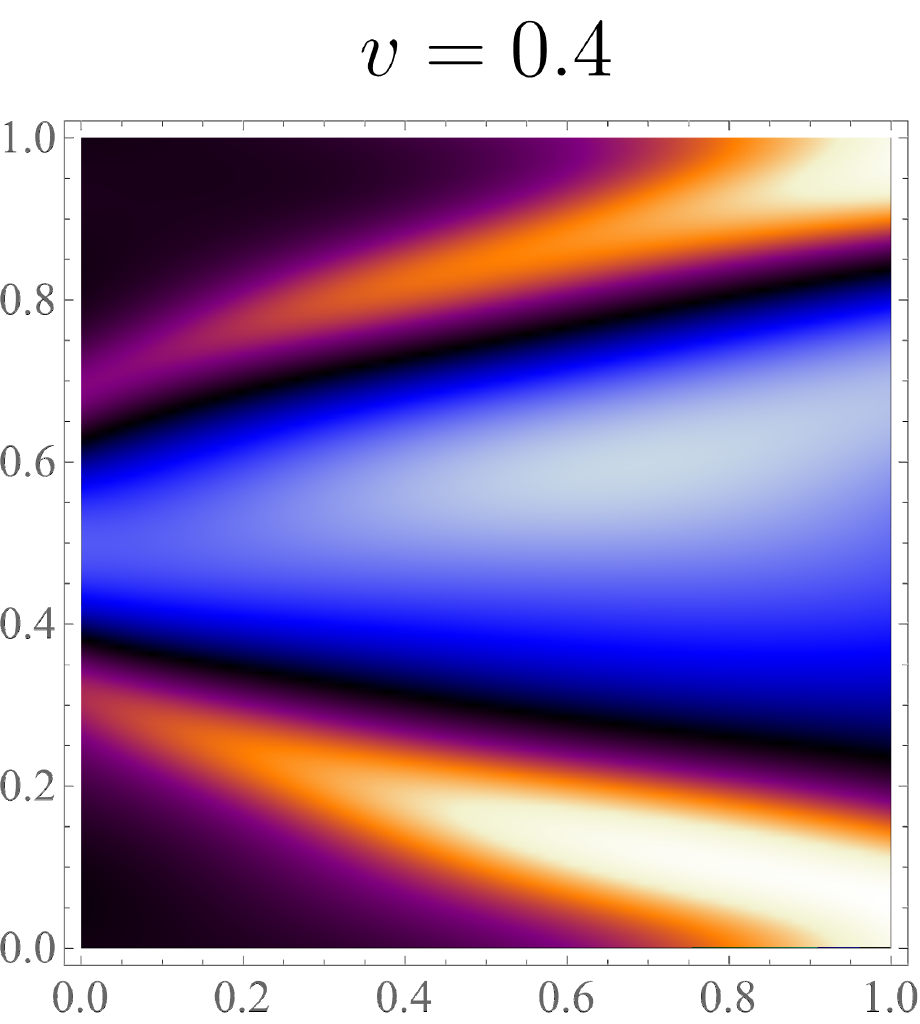} \\
     \includegraphics[width=0.27\linewidth, align=c]{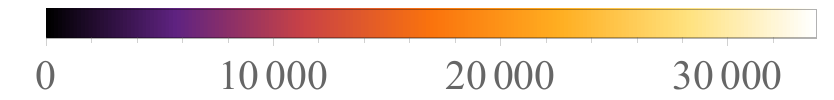} &
     \includegraphics[width=0.27\linewidth, align=c]{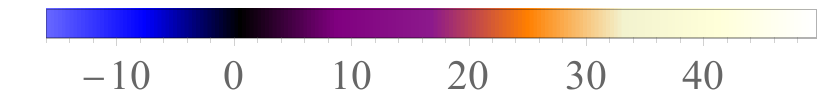} &
     \includegraphics[width=0.27\linewidth, align=c]{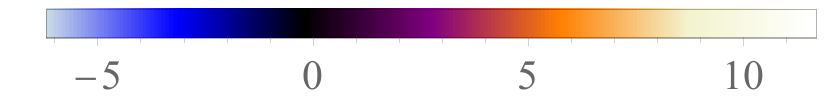} \\
     \includegraphics[width=0.27\linewidth, align=c]{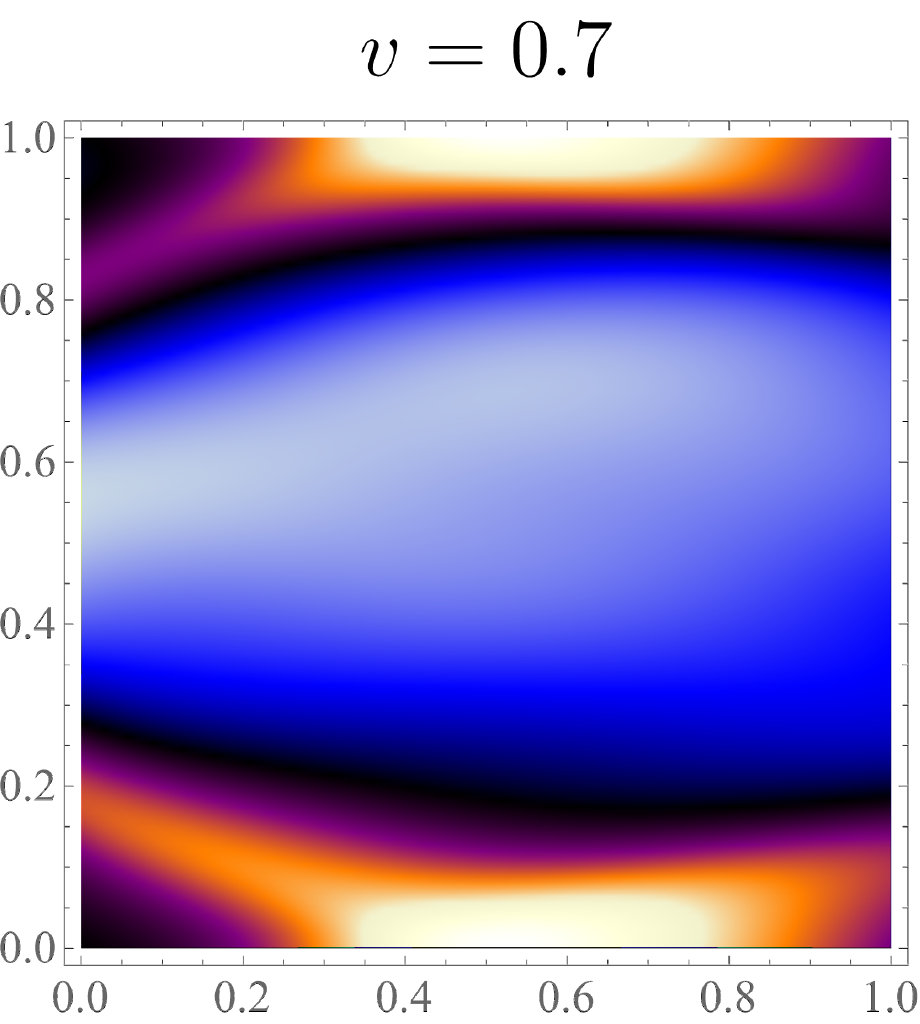} &
     \includegraphics[width=0.27\linewidth, align=c]{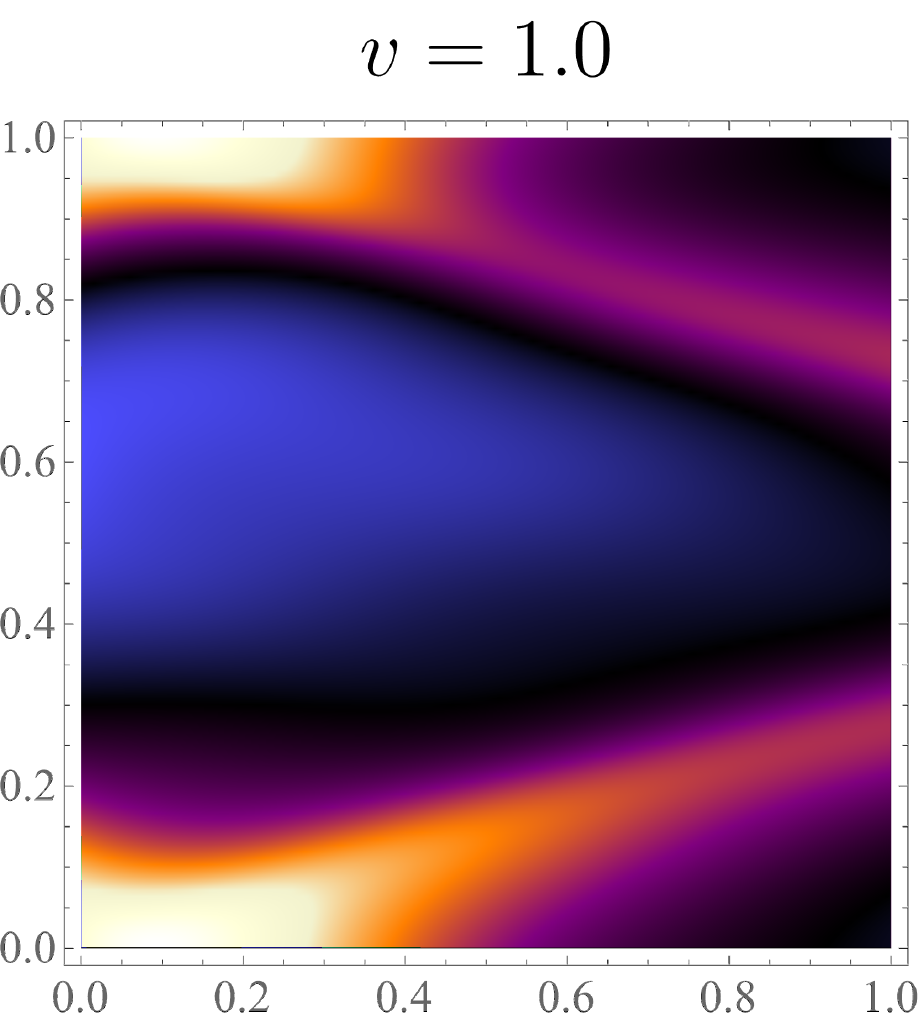} &
     \includegraphics[width=0.27\linewidth, align=c]{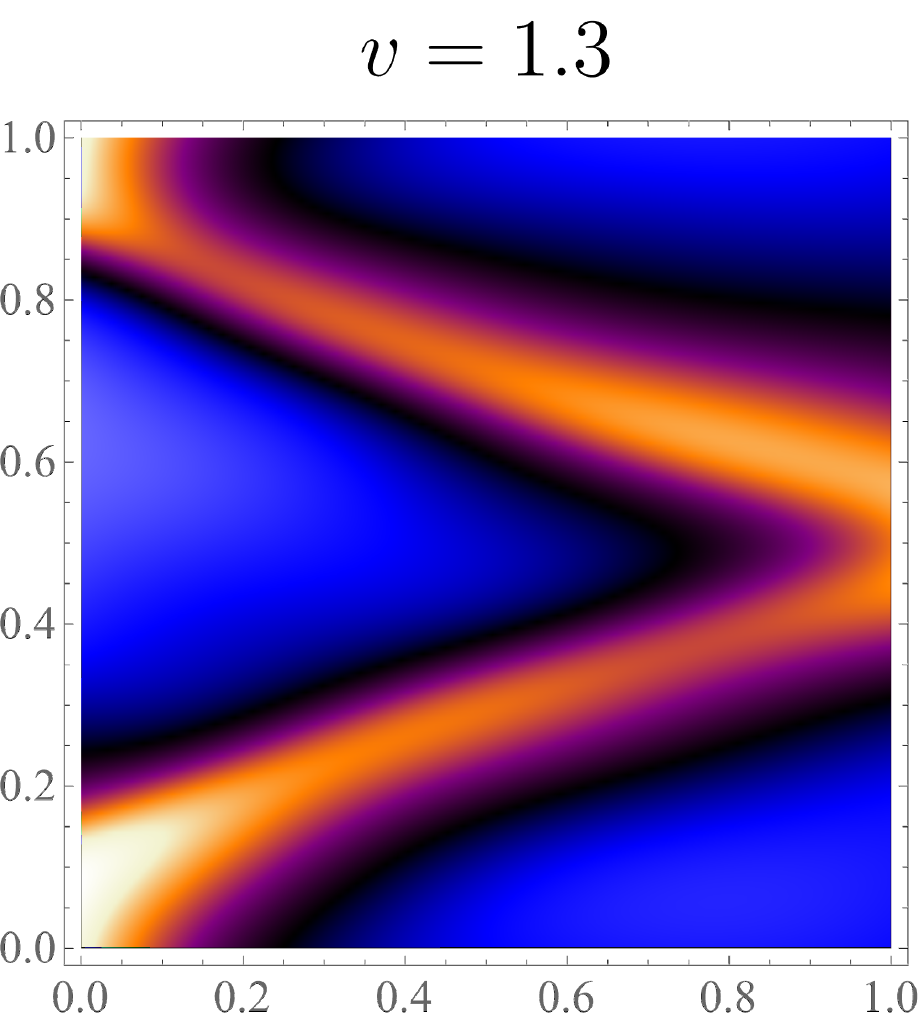} \\
     \includegraphics[width=0.27\linewidth, align=c]{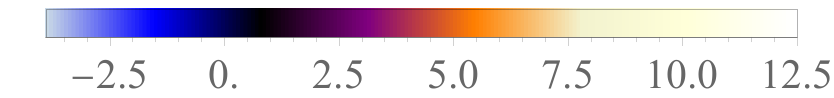} &
     \includegraphics[width=0.27\linewidth, align=c]{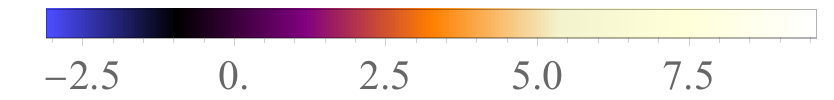} &
     \includegraphics[width=0.27\linewidth, align=c]{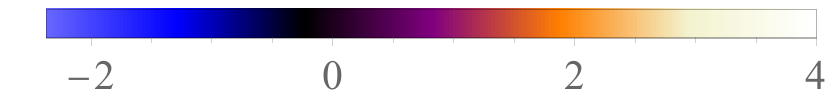}\\
     \includegraphics[width=0.27\linewidth, align=c]{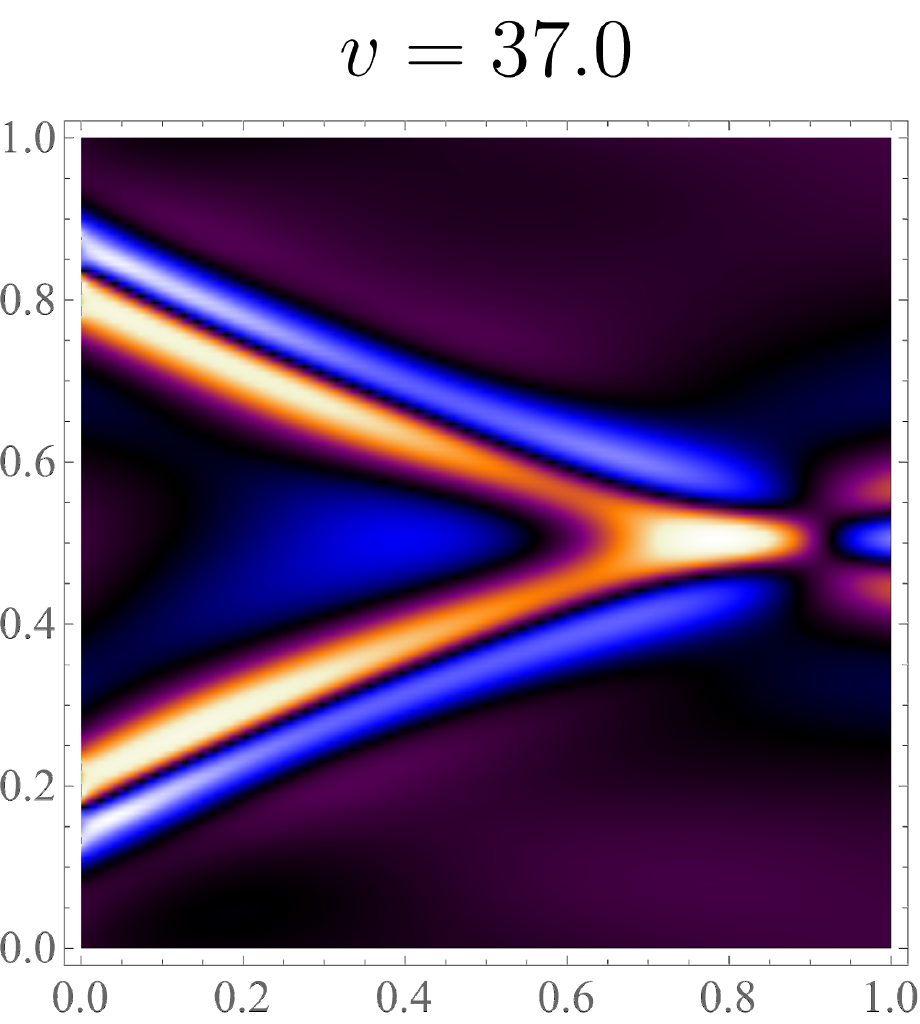} & 
     \includegraphics[width=0.27\linewidth, align=c]{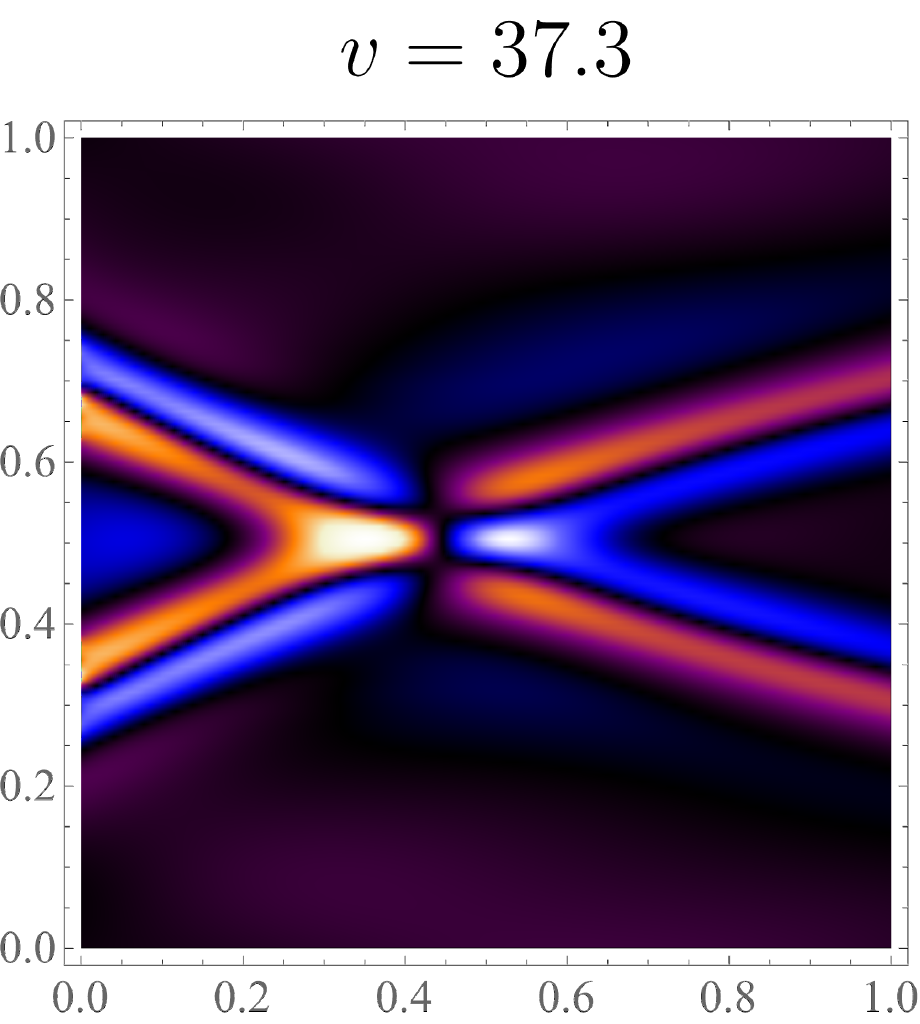} & 
     \includegraphics[width=0.27\linewidth, align=c]{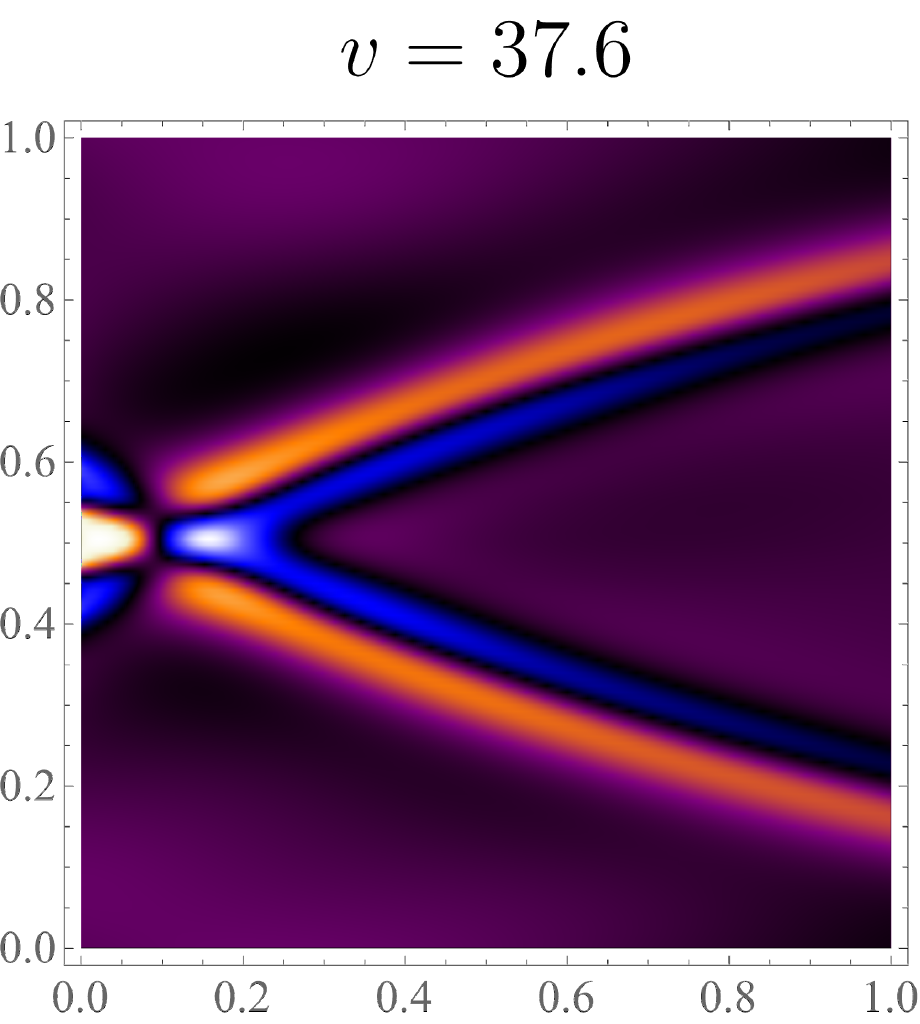} \\
     \includegraphics[width=0.27\linewidth, align=c]{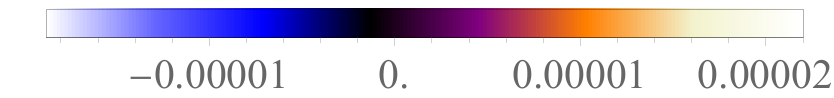} &
     \includegraphics[width=0.27\linewidth, align=c]{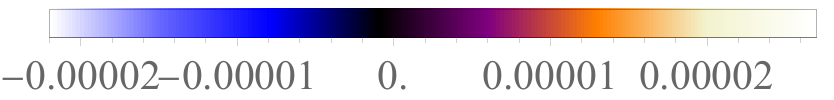} &
     \includegraphics[width=0.27\linewidth, align=c]{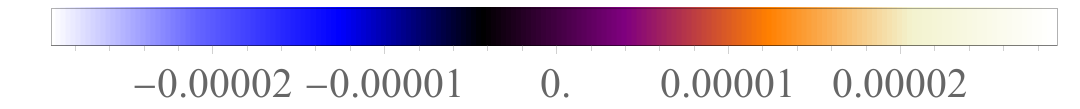}\\
     \includegraphics[width=0.27\linewidth, align=c]{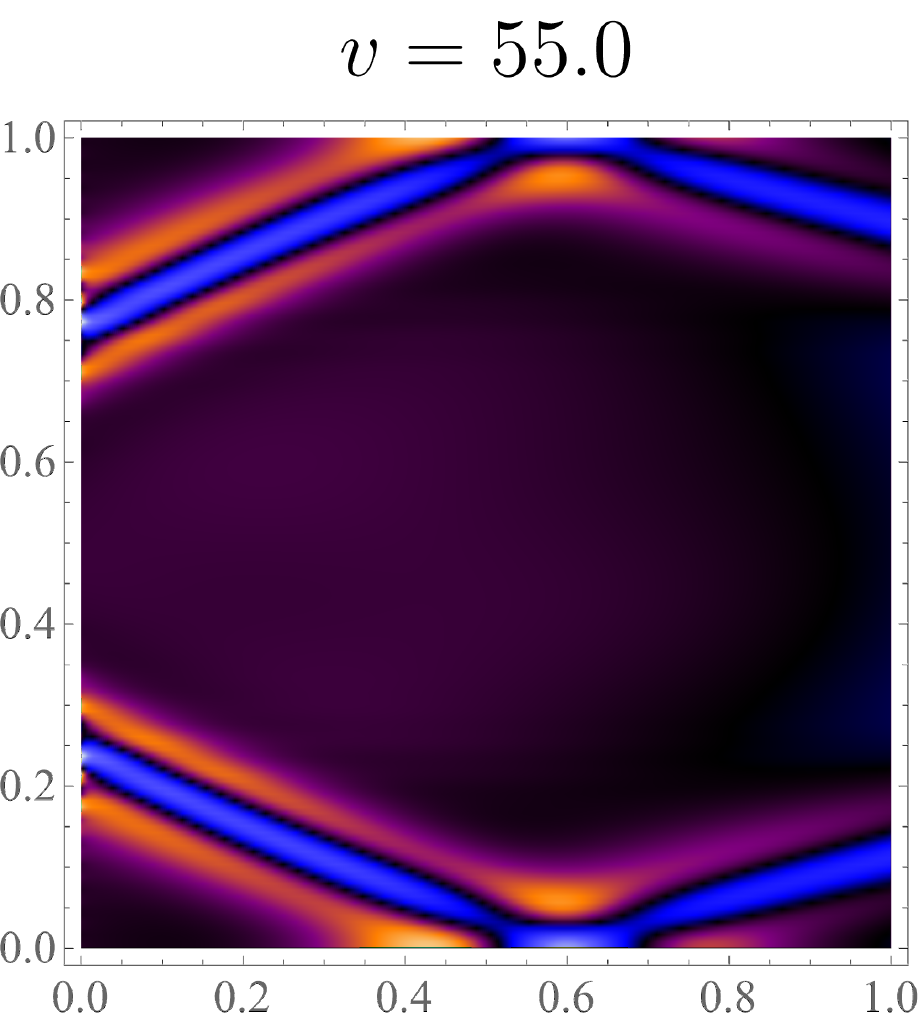} & 
     \includegraphics[width=0.27\linewidth, align=c]{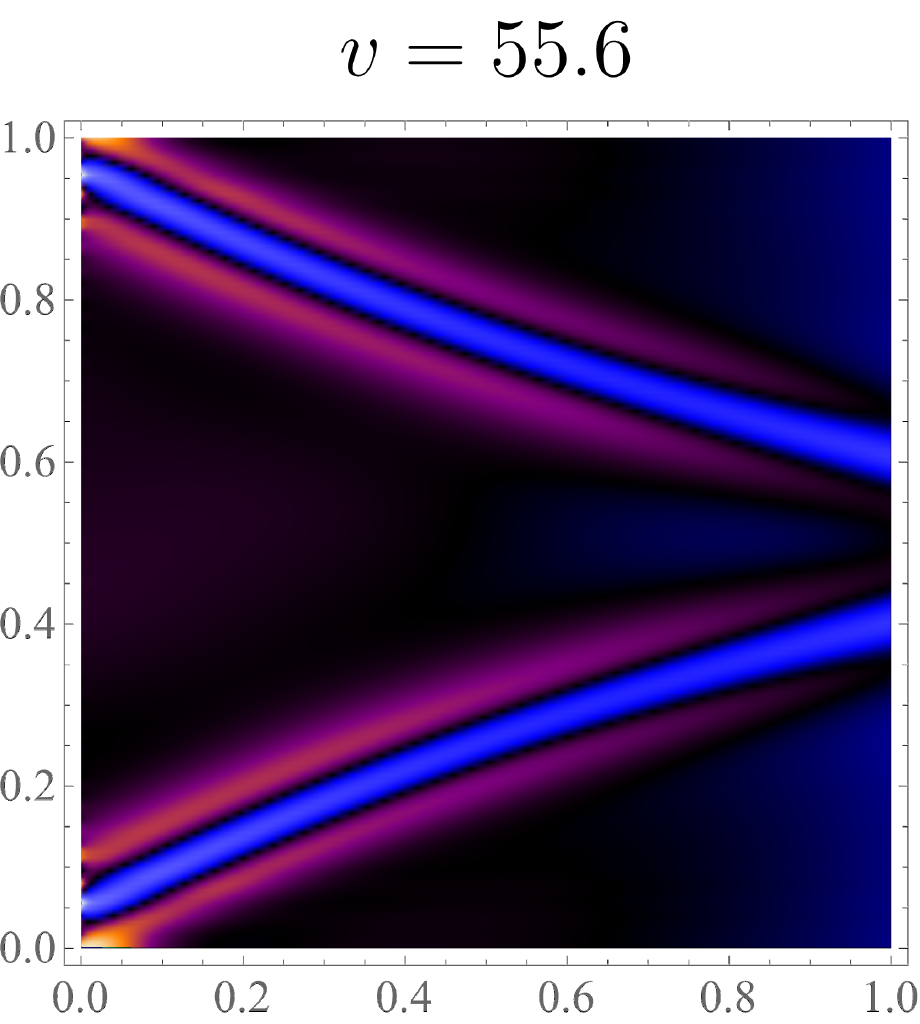} & 
     \includegraphics[width=0.27\linewidth, align=c]{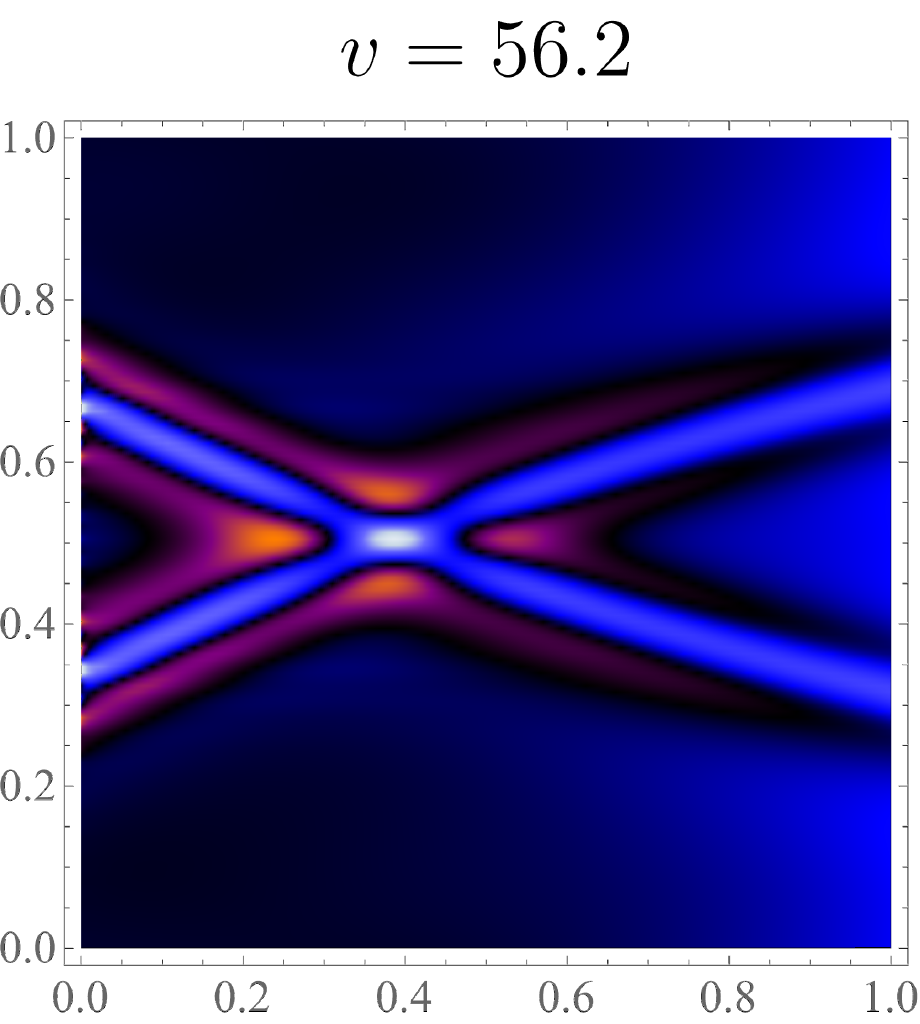}\\
     \includegraphics[width=0.27\linewidth, align=c]{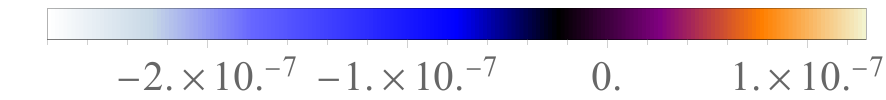} &
     \includegraphics[width=0.27\linewidth, align=c]{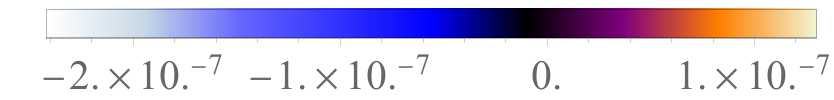} &
     \includegraphics[width=0.27\linewidth, align=c]{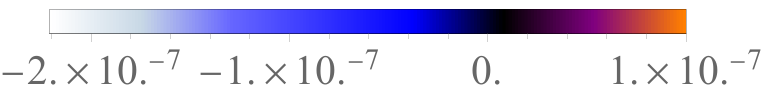}
     \end{tabular}
    \caption{Snapshots of the spatial profile of the Weyl scalar $(\ell^4 C_{abcd}C^{abcd}-12z^6)/z^{6}$ with the exact black brane value $12z^6$ subtracted, divided by the $z\to0$ scaling $z^6$. On each plot, the horizontal axis is $z$ and the vertical axis is $x/L$.}\label{fig:Kretschmann_small_black_hole}
\end{figure}

$V_3$ and $U_3$ provide insight into the behaviour at the boundary, but to study the behaviour in the bulk we use the Weyl scalar $C^2=C_{abcd}C^{abcd}$. This is related to the Kretschmann scalar $K=R_{abcd}R^{abcd}$ in AdS$_4$ by $C^2=K-24/\ell^2$, and the exact black brane solution has $C^2=12z^6/\ell^4$ in these coordinates. In Fig.~\ref{fig:Kretschmann_small_black_hole}, we show snapshots of the spatial profile of $(\ell^4C^2-12z^6)/z^6$ during the evolution. We observe the same qualitative behaviour as $V_3$ and $U_3$, now extending to $z$ as well as $x$.

\begin{figure}[t]
    \centering
    \begin{tabular}{@{}c@{}c@{}}
    \includegraphics[width=0.88\linewidth, align=c]{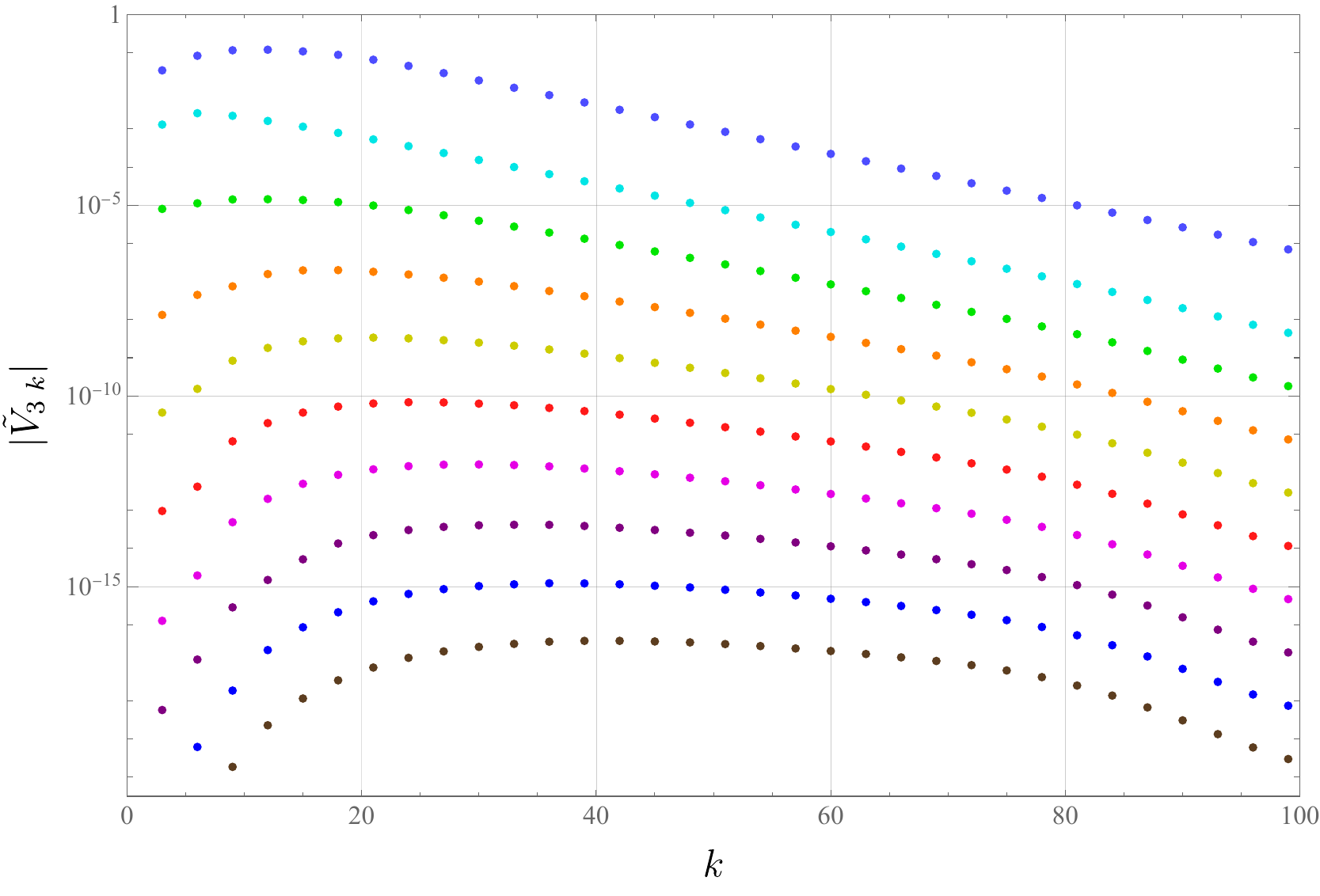} & \includegraphics[width=0.12\linewidth, align=c]{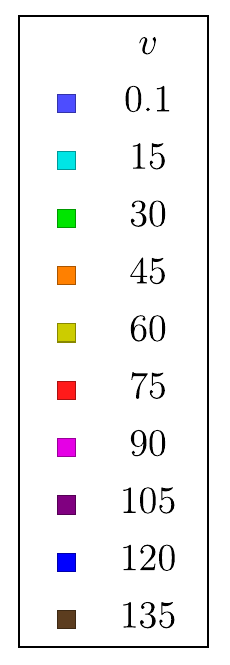}
    \end{tabular}
    \caption{Snapshots of the $V_3$ Fourier spectrum throughout the evolution. Each point represents the maximum value of the mode amplitude over a time window $\Delta v=0.1$ to hide the oscillatory behaviour.}\label{fig:V3_spectrum_snapshots}
\end{figure}

In Fig.~\ref{fig:V3_spectrum_snapshots}, we show snapshots of the Fourier spectrum of $V_3$ throughout the evolution. This shows the peak of the spectrum moving to higher wavenumber $k$ and broadening with time. It also shows just how fast the modes decay - a limitation of our method is that whilst there are QNMs that decay arbitrarily slowly at high $k$, the slow scaling of the decay rate with $k$ means that the resolvable modes in our numerical simulation all have relatively similar fast decay rates.

\begin{figure}[t]
    \centering
    \includegraphics[width=0.75\linewidth]{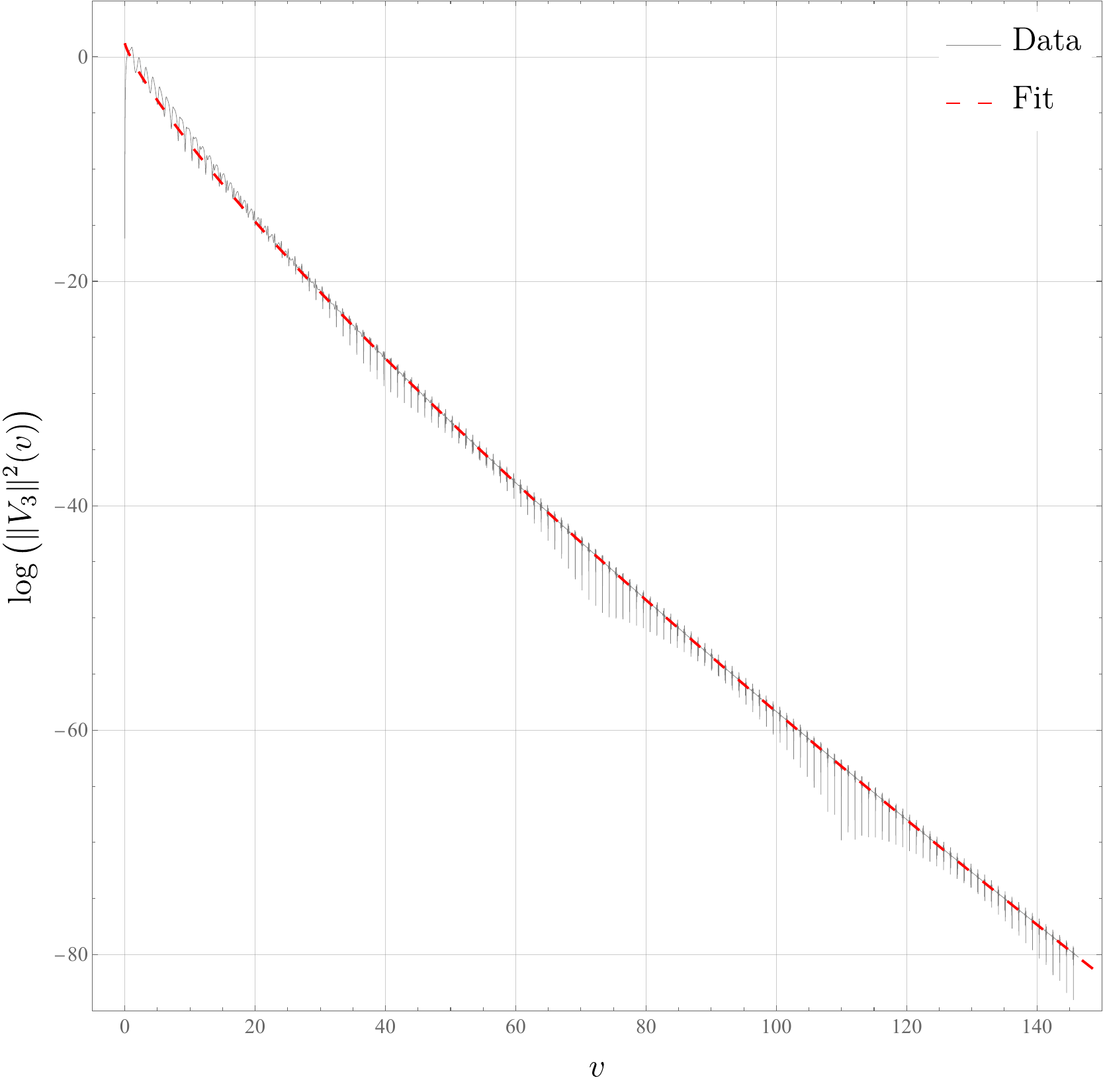}
    \caption{The decay of $\log{(\| V_3\|^2)}$ for the small $L=2\pi/3$ black hole is shown in grey along with a nonlinear fit to the form $-c_1 v^{\lambda}+c_2$, shown as a dashed red line, with fit parameters $c_1$, $c_2$, and $\lambda$. This is fitted using the data at times $v>90$ as it is the late time behaviour that we are predicting. The fitted exponent is $\lambda=0.821\pm0.01$, which is in agreement with the predicted value $\lambda=5/6$.}
    \label{fig:V3_prediction_plot}
\end{figure}

In Fig.~\ref{fig:V3_prediction_plot}, we explore the quantitative prediction for the late time behaviour. Recall from \eqref{eq:subexponential_late_time_prediction} the prediction of subexponential decay of the form $\log{(\| V_3\|^2(v))} \approx -c_1 v^{5/6}+\frac{5}{12}\log{(v)} + c_2$ with $c_1>0$, $c_2 \in \mathbb{R}$. We plot $\log{(\| V_3\|^2(v))}$ against $v$ as well as a nonlinear fit to the form $-c_1 v^{\lambda}+c_2$ with fit parameters $c_1$, $c_2$, and $\lambda$. The predicted value of $\lambda$ is $5/6$, and the fit returns $\lambda=0.821\pm0.01$ which lies within $2\%$ of the predicted value. This provides evidence in support of our prediction of subexponential decay. However, we were not able to accurately observe the $\frac{5}{12}\log{(v)}$ contribution to $\log{(\| V_3\|^2)}$. This may be because this contribution is considerably smaller than the leading subexponential term and would require a higher resolution solution with a longer runtime to observe. Equally, this could also indicate the absence of this behaviour in reality. The resolution of this question represents an avenue for future work.

\newpage

\subsection{A large black hole: $L=2\pi$}

We now explore the case of a black hole with $L=2\pi$. This is large in the sense that $L>2\pi/k_{\mathrm{max}}$, and the longest wavelength mode $k=1$ has a significantly smaller QNM decay rate than a large number of $k>1$ modes - it is only once $k$ reaches several orders of magnitude higher that the decay rates return to below that of the $k=1$ mode. To illustrate this, the decay rates of the QNMs with respect to $k$ are shown for $L=2\pi$ in Fig.~\ref{fig:AdS4_L_comparison_decay_rates}. On the other hand, this is not a large black hole in the hydrodynamics setting. The longest allowed wavelength $\lambda=2\pi$ is still too short for the spectrum of decay rates at low $k$ to show the diffusive $k^2$ scaling associated with long wavelength low frequency hydrodynamics - the long-lived $k=1$ mode is the only remnant. Even for very large black holes with $L\gg 2\pi/k_{\mathrm{max}}$, the very fact that there is a maximum wavelength $\lambda=L$ mode but no minimum wavelength mode means that eventually (after an exceedingly long time) the large $k$ tail will dominate the behaviour. In this study, the choice $L=2\pi$ is a compromise where we can observe the coexistence of a long-lived $k=1$ mode alongside a long spectral tail (in fact, $k=2$ and $k=3$ are also below $k_{\mathrm{max}}$ but not as significantly as $k=1$).

For initial data, we choose
\begin{subequations}
\begin{align}
    Q_1(0,z,x) &= 0.01 \left[  \frac{\cos{x}}{1.1+\cos{x}} + \sin^3{x} -  \sum_{m=-2}^{m=2} a_m e^{{\rm i}\,m\,x}  \right]\\
    V_3(0,x) &= 0\\
    U_3(0,x) &= 0
\end{align}
\end{subequations}
\noindent where the coefficients $a_m$ are chosen to exactly remove the $-2\leq m \leq 2$ modes from the $Q_1$ profile. These are $a_0=1-11/\sqrt{21}$, $a_1=(11\sqrt{21}-21)(11/210) - (3/8)i$, $a_3=121/50-781/(50 \sqrt{21})$, and the rest are determined by $a_{-m}=a_m^*$. This choice is unusual, but we choose it in an attempt to suppress the long-lived $k=1,2$ modes during the early time evolution as to not mask the higher $k$ behaviour. They will only be switched on through nonlinear interactions, and so will be relatively small initially. However, we do nonetheless expect these modes to have an era of dominance due to their low decay rates - especially the $k=1$ mode.

We use the same numerical setup as the $L=2\pi/3$ solution, but use $256$ collocation points for the $x$ grid instead of the previous $128$ points. We use a timestep $\Delta v=0.0005$ and evolve up to $v=165$.

\begin{figure}[t]
\begin{tabular}{@{}c@{}c@{}c@{}c@{}}
     \includegraphics[width=0.43\linewidth, align=c]{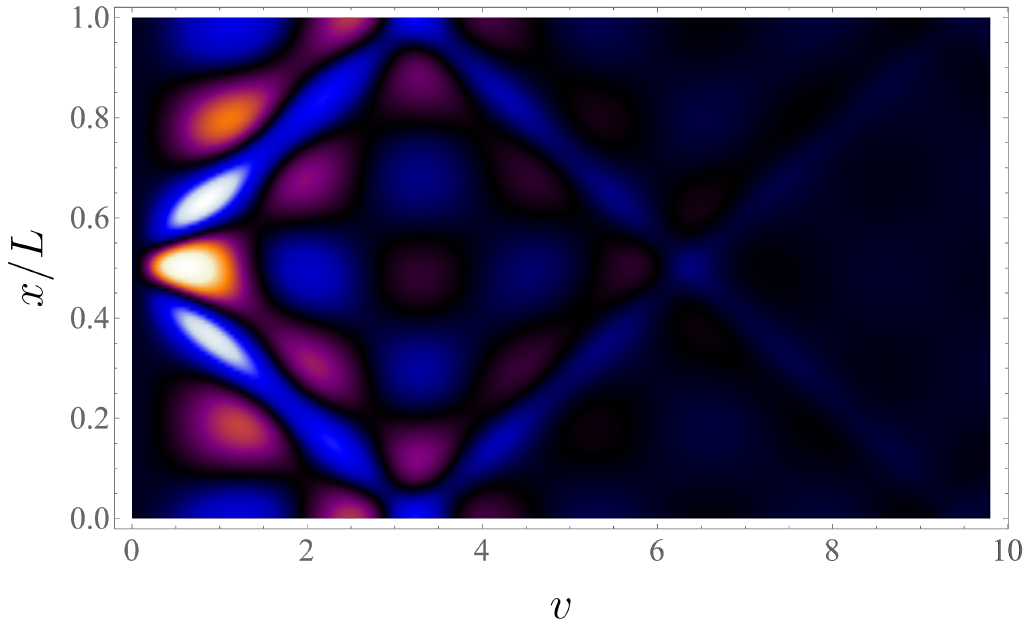} & \includegraphics[width=0.05\linewidth, align=c]{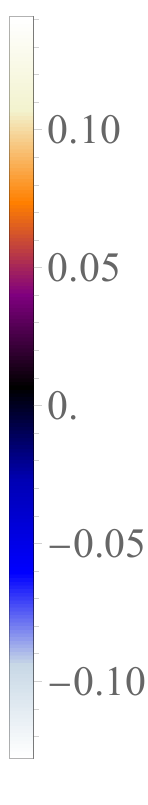} &
     \includegraphics[width=0.43\linewidth, align=c]{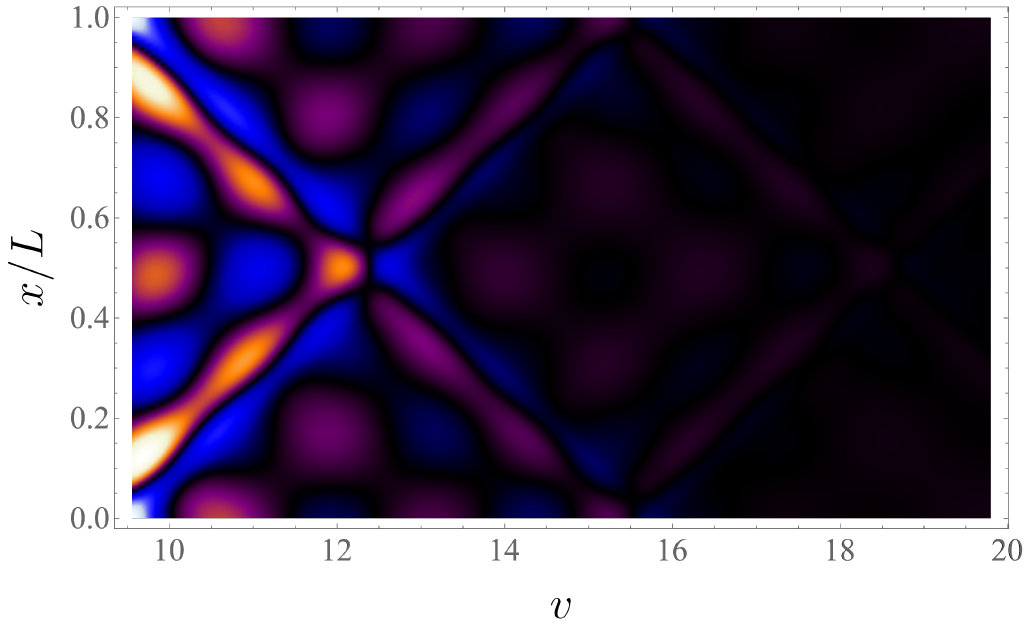} & \includegraphics[width=0.055\linewidth, align=c]{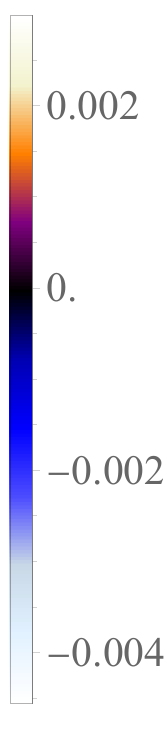}\\
     \includegraphics[width=0.43\linewidth, align=c]{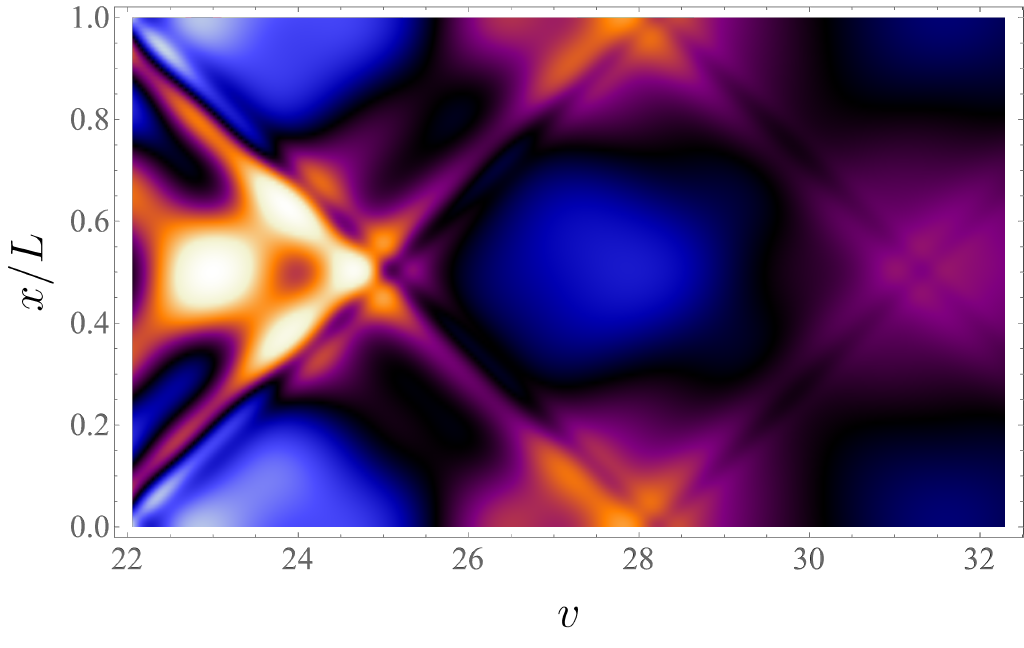} & \includegraphics[width=0.06\linewidth, align=c]{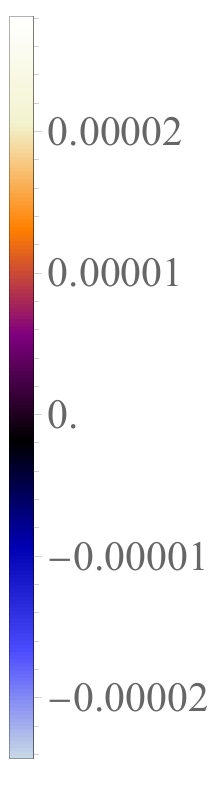} &
     \includegraphics[width=0.43\linewidth, align=c]{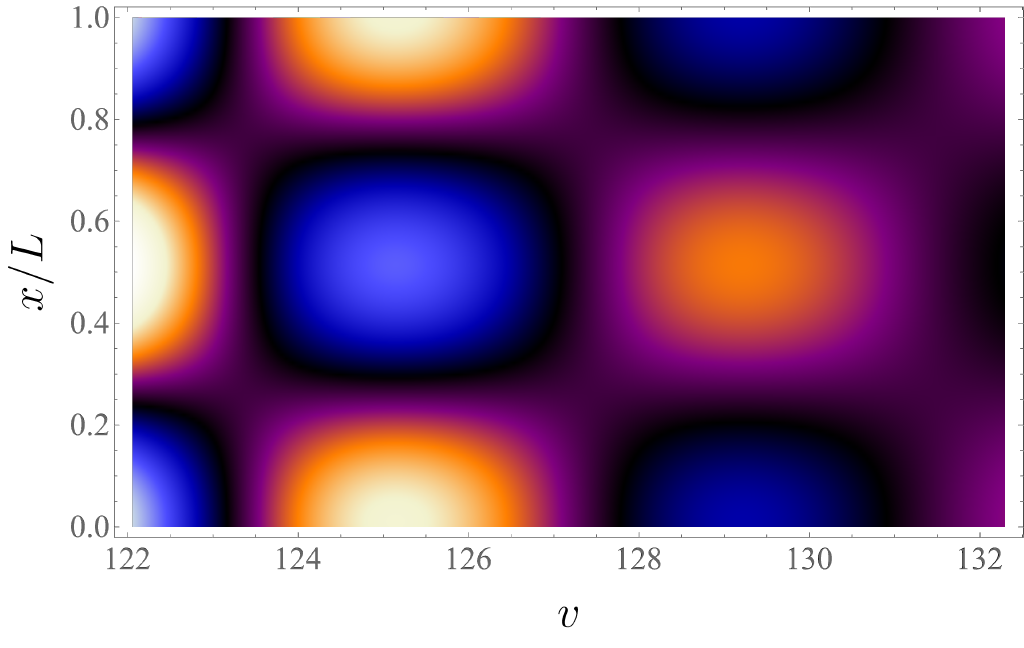} & \includegraphics[width=0.081\linewidth, align=c]{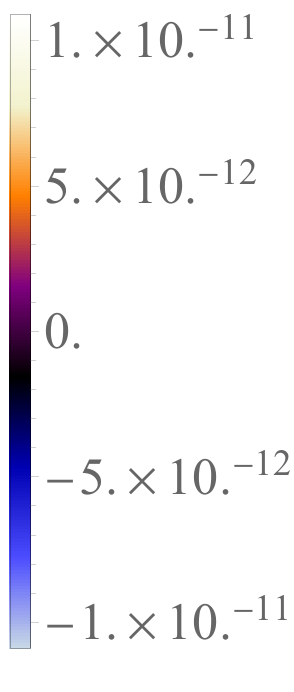}
\end{tabular}
    \caption{Time evolution of the boundary energy density $V_3$ during four time intervals. The $k=1$ mode, switched on through nonlinear interactions, outlasts the other resolvable modes. A longer runtime with a higher resolution would be required to observe the regime change when the high wavenumber tail begins to dominate the behaviour.}\label{fig:V3_large_black_hole_plot}
\end{figure}

\begin{figure}[t]
\centering
\begin{tabular}{ccc}
    \includegraphics[width=0.28\linewidth, align=c]{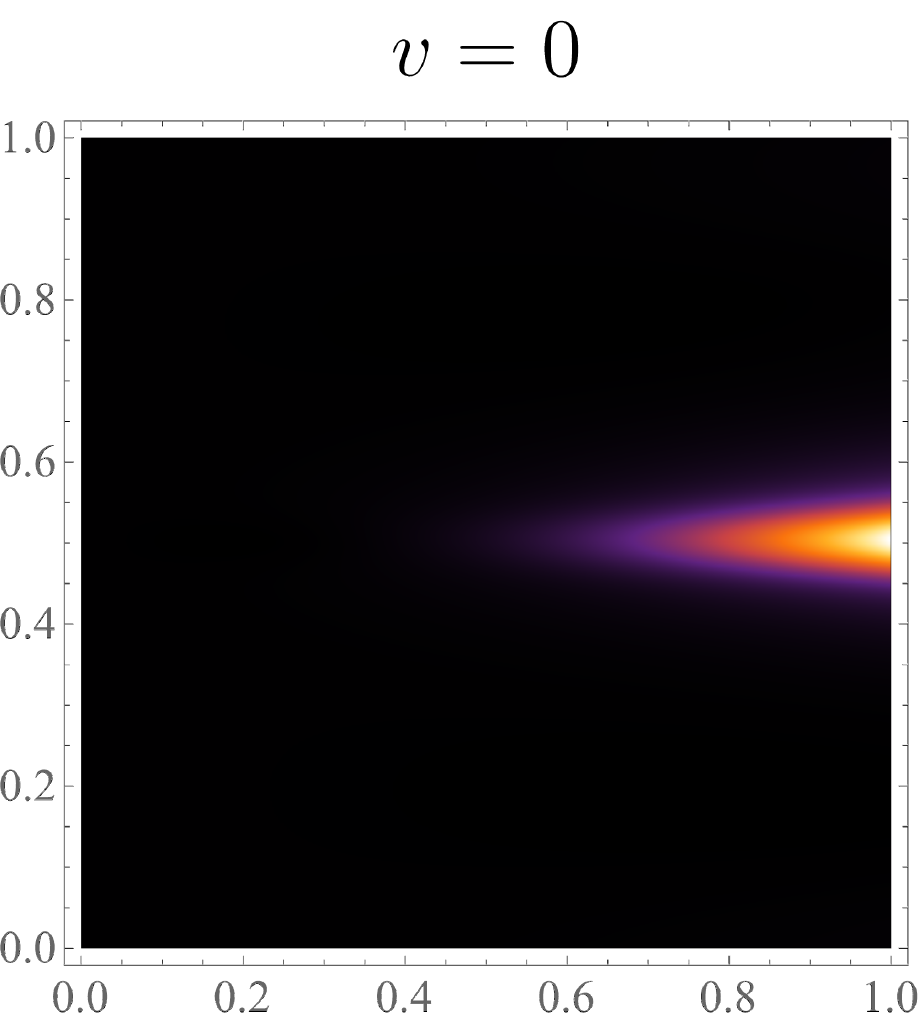} &
     \includegraphics[width=0.28\linewidth, align=c]{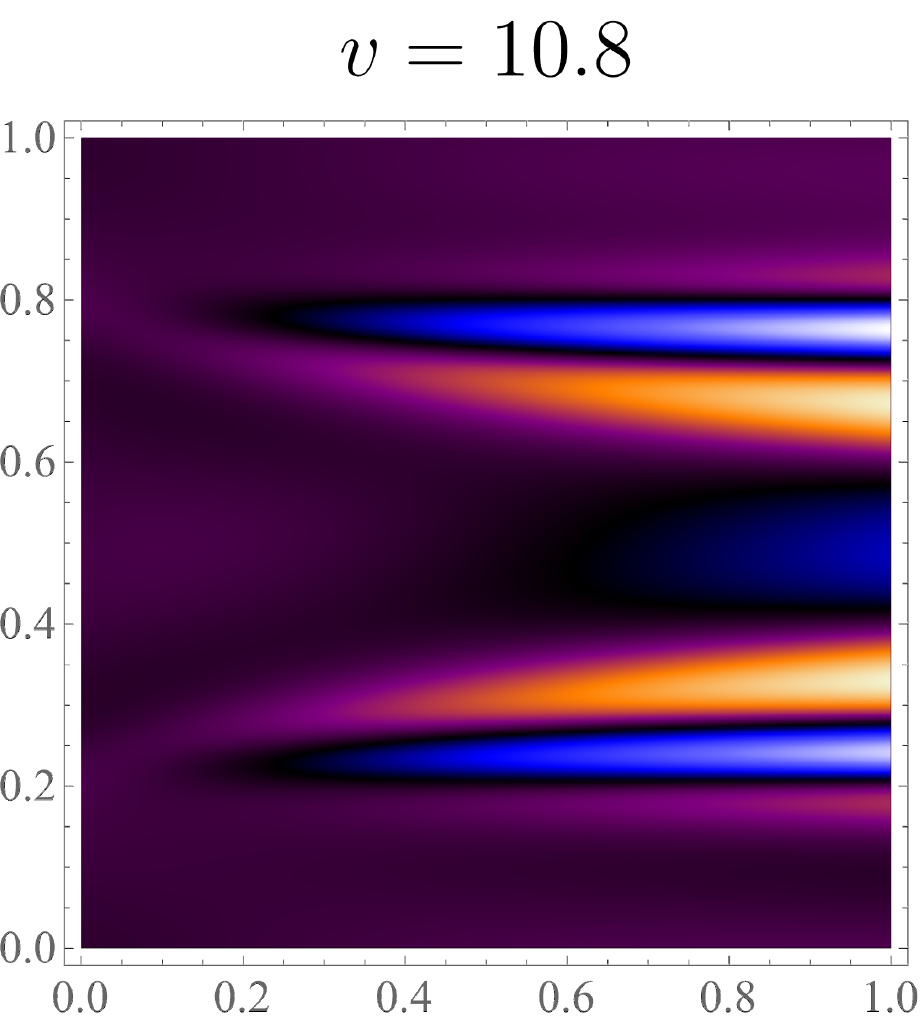} &
     \includegraphics[width=0.28\linewidth, align=c]{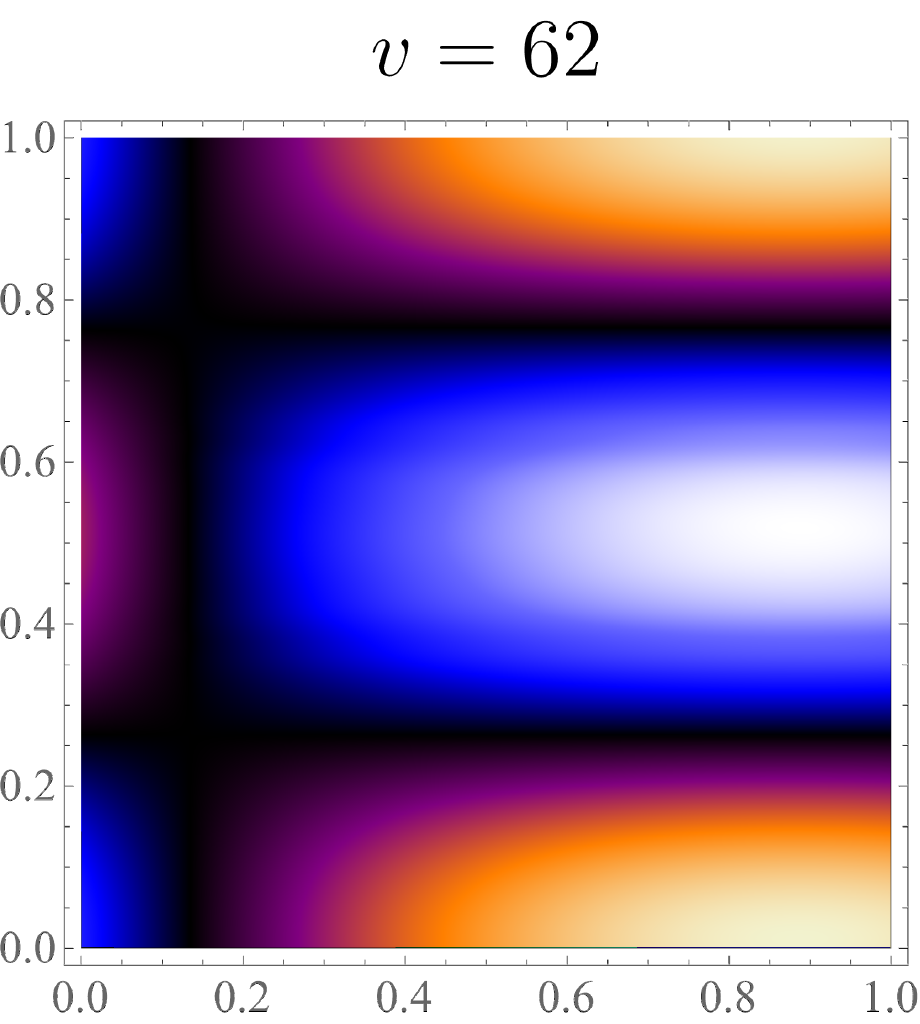} \\
     \includegraphics[width=0.28\linewidth, align=c]{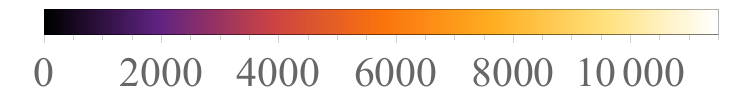} &
     \includegraphics[width=0.28\linewidth, align=c]{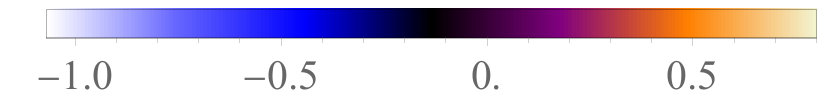} &
     \includegraphics[width=0.28\linewidth, align=c]{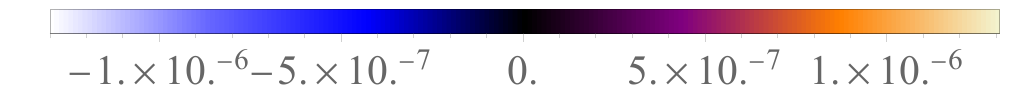}
     \end{tabular}
    \caption{Snapshots of the Weyl scalar $(\ell^4C^2-12z^6)/z^6$ with the exact black brane value subtracted and divided by the $z\to0$ scaling. On each plot, the horizontal axis is $z$ and the vertical axis is $x/L$. The long-lived $k=1$ mode outlasts the other resolvable modes and is prominent at late times.}\label{fig:Weyl_scalar_large_black_hole_AdS4}
\end{figure}

The evolution of $V_3$ is shown in Fig.~\ref{fig:V3_large_black_hole_plot} during different time intervals. Initially it appears similar to the small black hole case, with short wavelength oscillations and a narrowing wavepacket. However, as time progresses the $k=1$ mode becomes increasingly visible as the faster decaying higher $k$ modes become smaller. By the end of the runtime, only the $k=1$ mode is visible. This behaviour is also reflected in the evolution of the Weyl scalar, shown in Fig.~\ref{fig:Weyl_scalar_large_black_hole_AdS4}. 

\begin{figure}[t]
    \centering
    \begin{tabular}{@{}c@{}c@{}}
    \includegraphics[width=0.88\linewidth, align=c]{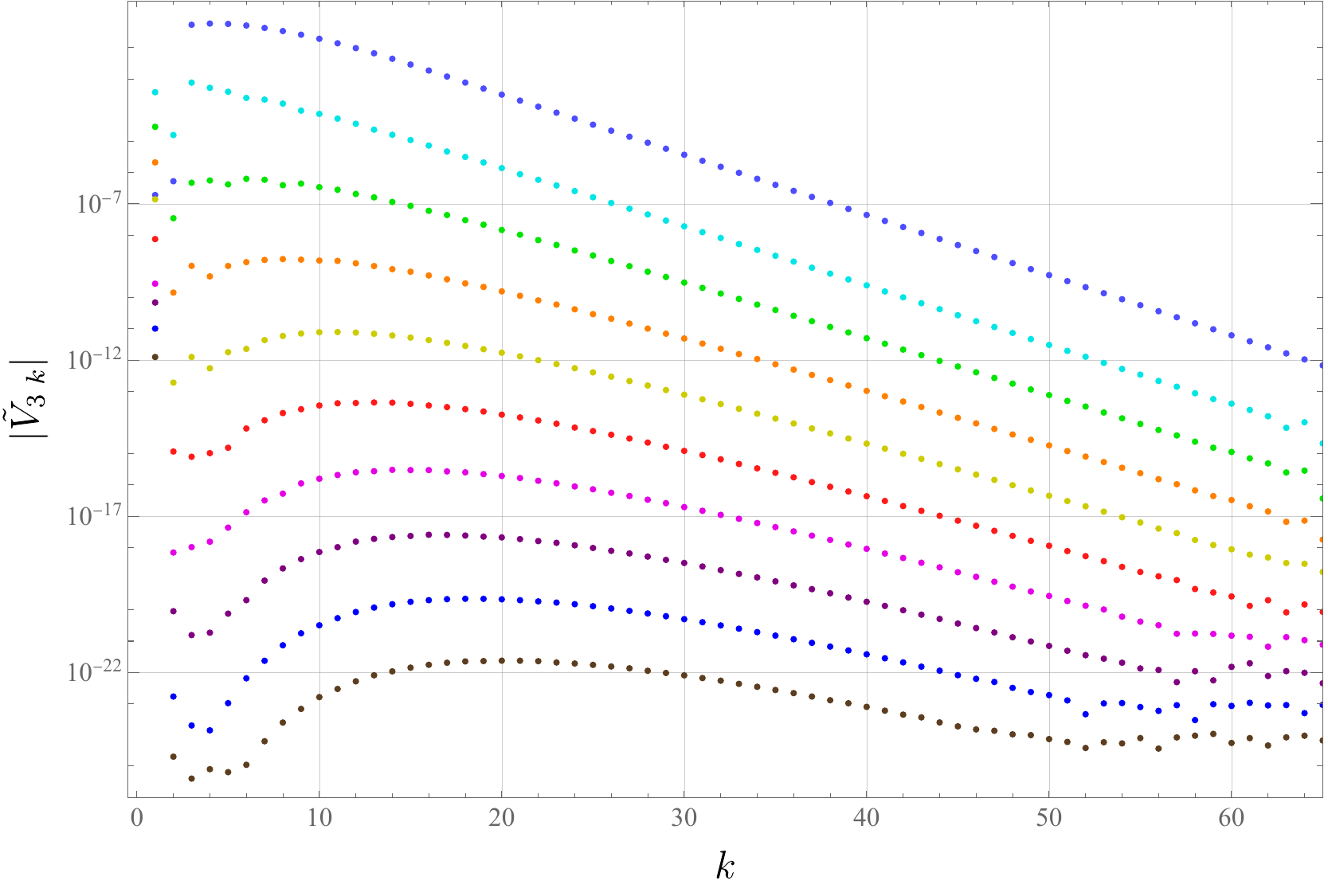} & \includegraphics[width=0.12\linewidth, align=c]{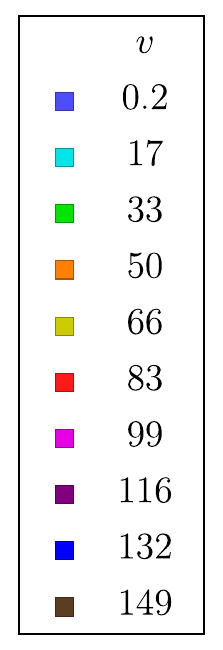}
    \end{tabular}
    \caption{Snapshots of the $V_3$ Fourier spectrum throughout the evolution. Each point represents the maximum value of the mode amplitude over a time window $\Delta v=0.1$ to hide the oscillatory behaviour. The long-lived $k=1$ mode is relatively small initially, but outlasts the other resolvable modes. Distortion of the spectrum at large $k$ is due to accumulating numerical errors.}\label{fig:V3_spectrum_snapshots_large_black_hole}
\end{figure}

We can see this by looking directly at snapshots of the spectrum of $V_3$, shown in Fig.~\ref{fig:V3_spectrum_snapshots_large_black_hole}. As intended from the choice of initial data, the $k=1$ mode is initially relatively small compared to the higher modes after being switched on by nonlinear interactions. By $v\approx20$ though, it is already the dominant mode in the spectrum and this remains the case for the rest of the runtime. Even with longer runtimes, this setup would not see the late time regime change because the next mode with a slower decay rate than the $k=1$ mode has a wavelength significantly smaller than the grid spacing used here. Furthermore, as can be seen in the spectrum plot, by the end of the runtime the higher modes are already so small that they are reaching the threshold of machine precision and numerical errors are becoming significant.

\begin{figure}[t]
    \centering
    \includegraphics[width=0.75\linewidth]{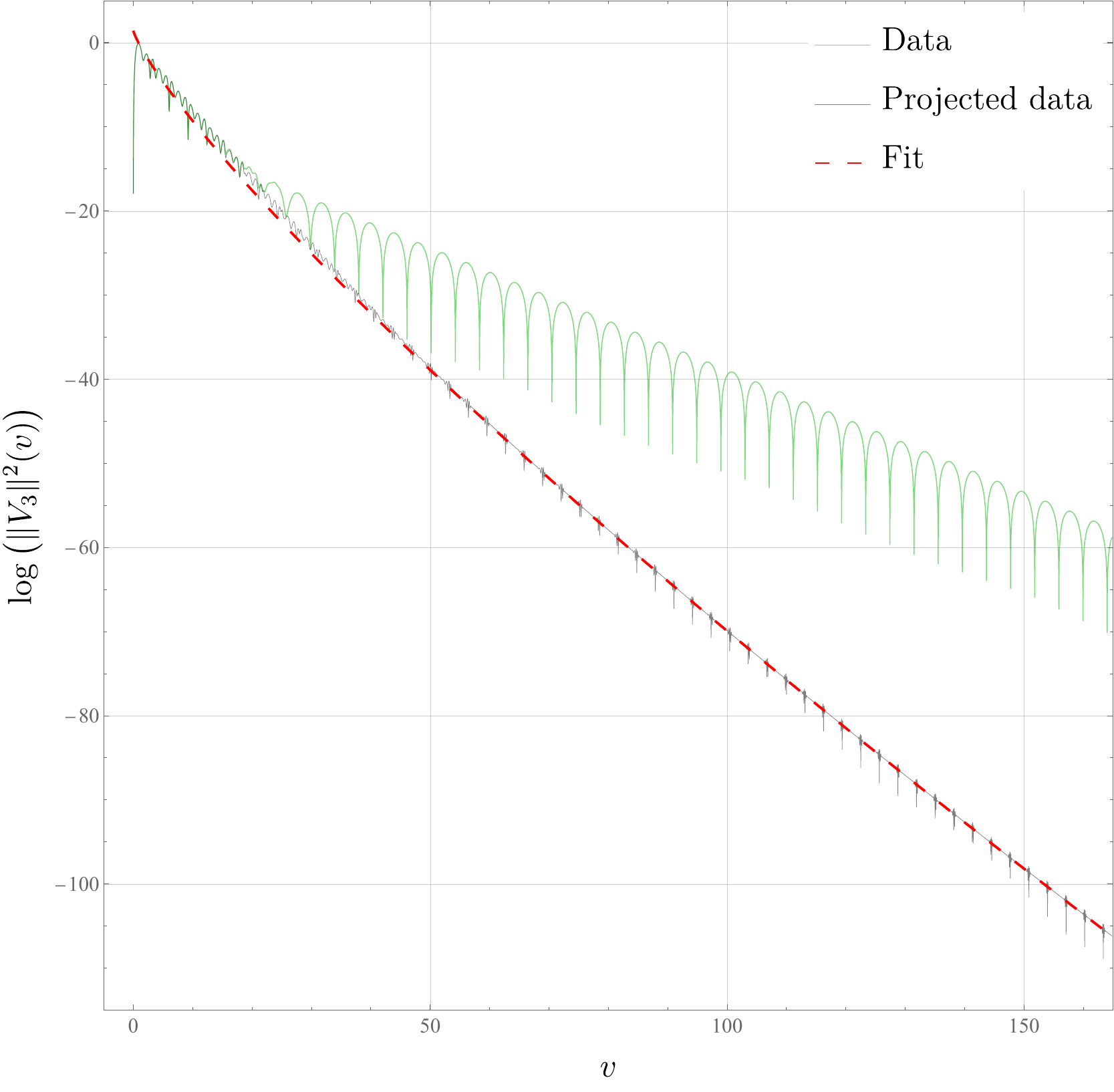}
    \caption{The decay of $\log{(\| V_3\|^2)}$ for the $L=2\pi$ black hole is shown in green, and the grey line shows $\log{(\| V_3\|^2)}$ after artificially projecting out the $k=1$ mode from $V_3$. The dashed red line shows the fit of the projected data to the predicted form $-c_1 v^{\lambda}+c_2$. This was fitted on the window $v>110$. The fitted exponent is $\lambda=0.823\pm0.01$. }\label{fig:V3_prediction_plot_large_black_hole}
\end{figure}

Despite the dominance of the $k=1$ mode, we can still see the expected behaviour in the rest of the spectrum. The peak of the spectrum (excluding $k=1$) moves to higher $k$ with time and exhibits broadening. We can go one step further, and test the quantitative prediction of subexponential decay by projecting out the $k=1$ mode from $V_3$ and then calculating $\log{(\|V_3\|^2)}$. This is shown in Fig.~\ref{fig:V3_prediction_plot_large_black_hole}, which also includes $\log{(\|V_3\|^2)}$ \textit{without} projecting out the $k=1$ mode. At $v\approx20$ we see that the $k=1$ mode begins to dominate, as seen previously in the spectrum plot. The line for the projected version however appears qualitatively similar to the $L=2\pi/3$ case. We can perform a nonlinear fit to this projected data, fitting to the predicted form $-c_1 v^{\lambda}+c_2$ for fit parameters $c_1$, $c_2$, and $\lambda$. This produces $\lambda=0.823\pm0.01$ which is again close to the predicted value of $5/6$. This fitted line is also shown in Fig.~\ref{fig:V3_prediction_plot_large_black_hole}.

This solution does not reach the late time regime, as that would be the regime in which the high $k$ modes dominate over all of the lower $k$ modes, including the slowest decaying $k=2\pi/L$ mode. The presence of this subexponential decay in the projected data before reaching that regime is likely due to the effective decoupling of the modes as they have already become very small, placing us in the regime well-approximated by the linearised equations. However, we do still expect that the late time regime would eventually be reached with a longer runtime and higher resolution. At that point, we expect that this subexponential scaling would become apparent in the data without projecting out any modes.

\section{Void Formation and Knudsen-Like Transport in AdS$_4$}
The late-time behaviour uncovered in our planar AdS$_4$ evolution suggests a qualitative change in the mechanism governing relaxation. At early times the dynamics are well described by the usual hydrodynamic sector: long-wavelength modes dominate, diffusion controls the return to equilibrium, and the decay of perturbations exhibits the familiar quasinormal damping characteristic of a black hole background. This is the regime typically associated with the thermal phase in the dual CFT.

At sufficiently late times, however, the situation changes dramatically. The evolution becomes dominated by a sector of high-momentum modes whose damping rates scale as
\begin{equation}
\operatorname{Im}\omega_{k\,n} \sim k^{-1/5},
\end{equation}
leading to an accumulation of infinitely many quasinormal modes arbitrarily close to the real axis. These modes propagate with nearly constant group velocity and behave to good approximation as weakly damped, nearly ballistic excitations, resembling normal modes of pure AdS. In spacetime diagrams this transition is accompanied by the formation of extended regions where the scalar amplitude becomes exponentially small-dynamically generated voids produced by destructive interference among large-$k$ wave packets. Within these voids the dynamics is effectively ballistic, with excitations travelling along null rays and interacting only at the sharp fronts that define the void boundaries.

Although these voids arise purely from deterministic wave propagation and do not reflect any underlying stochasticity, their spectral effect mirrors that of rare regions in disordered systems: they suppress the effective contribution of low-frequency modes to late-time dynamics and enhance the role of high-$k$ modes. This reshaping of spectral weight forces the system into a non-hydrodynamic regime where relaxation is governed by the high-frequency tail and exhibits stretched-exponential decay. In this sense, the voids act analogously to kinetic pores in Knudsen diffusion, isolating the ballistic carriers from the strongly damped, hydrodynamic sector. A closely related phenomenon appears in a different physical setting in recent work on many-body systems \cite{McCulloch:2025fzk}, where stretched exponentials arise from rare-region (“void’’) effects in disordered or effectively disordered media.

The resulting spectral flow has a natural interpretation in terms of AdS/CFT. Although the bulk geometry remains that of a planar AdS black hole throughout the evolution, the late-time dynamics is effectively insensitive to the near-horizon region. The high-$k$ modes responsible for the stretched exponential primarily probe the ultraviolet portion of the geometry, where the metric is nearly AdS. As a result, the CFT dynamics crosses over from a thermal, diffusion-dominated regime to one dominated by long-lived, vacuum-like excitations. This transition is not a genuine thermodynamic phase change - nothing in the bulk geometry undergoes a Hawking-Page-type instability - but the spectral character of the excitations changes in a way that is structurally analogous to the distinction underlying the Hawking-Page transition.

Viewed through this lens, the phenomenon we observe may be understood as a spectral crossover: the system begins in the quasinormal-mode-dominated, thermal regime, but as time evolves and the high-momentum modes become dominant, the late-time response increasingly resembles that of pure AdS. The analogy is not thermodynamic but analytic: the same distinction between hydrodynamic/diffusive relaxation and ballistic, normal-mode-like propagation that characterises the two sides of the Hawking-Page transition is reproduced here dynamically, as a consequence of the spectral flow induced by the real-analytic initial data.

This perspective unifies several elements of our findings: the void formation, the emergence of Knudsen-like transport in the CFT, the accumulation of quasinormal modes near the real axis, and the resulting stretched-exponential decay. All of them can be understood as consequences of a common underlying mechanism - the dominance of the high-frequency tail and the corresponding return of effectively vacuum-like propagation - which gives rise to a late-time dynamical regime that is qualitatively distinct from the early, hydrodynamic evolution. The analogy to the Hawking-Page transition therefore serves as a useful interpretive tool: it highlights the way in which the spectral structure, rather than the geometry itself, controls the effective phase of the dynamics.

\section{Discussion}
The results presented in this paper provide evidence for a late time subexponential decay of perturbations of toroidal AdS-Schwarzschild black holes in 4D, within the regularity class of real-analytic initial data. By studying the quasinormal mode spectrum, we expect that this regime eventually occurs for any horizon circumference, including for very large black holes usually studied in the context of the fluid/gravity correspondence which feature a long wavelength diffusive sector of the spectrum. This is predicted to occur at very late times, after the gradient expansion ceases to accurately describe the increasingly short length-scale behaviour.

The first case study explored a small black hole \textit{without} the hydrodynamic sector in order to highlight the time dependence that arises due to the tail of high momentum modes. We observed subexponential decay of the predicted form $\exp{(-c t^{5/6})}$ for $c>0$. The second case study looked at a larger black hole that featured three long-lived low wavelength modes -  remnants of the hydrodynamic sector. The results of this study are consistent with the picture that there will be a regime of subexponential decay of the form $\exp{(-c t^{5/6})}$ after the longest wavelength mode has decayed to be sufficiently small - smaller than the short wavelength modes that decay even more slowly. Whilst this regime was not reached in the solution presented here due to technical limitations, the observed behaviour is suggestive that this will eventually be the case.

This paper leaves open several questions for future exploration. The simulations presented here are only of black holes with a relatively small circumference. These are too small to have a large long-wavelength sector of the spectrum that yields \textit{bona fide} hydrodynamics in the boundary QFT. This choice not to simulate larger black holes is because these hydrodynamic modes, present for large black holes, can have exceedingly long lifetimes - the decay rate of the longest-lived mode scaling with $1/L^2$ for horizon circumference $L$. In contrast, the decay rates for the short wavelength modes approach zero very slowly, only scaling with $1/k^{1/5}$ for wavenumber $k$. This means waiting a very long time for the longest wavelength mode to sink to below the scale of the slower decaying modes at high $k$. The combination of the requirement of long runtimes to reach this regime, combined with the high resolution discretisation needed to resolve the high $k$ modes, means that this is beyond the scope of the methods used in this paper. Future work might hope to overcome these current limitations.

The numerical method for time evolution used here is fundamentally only able to evolve up to a finite future time. This means that any prediction of behaviour that appears at asymptotically late times can only be suggested by numerical evidence. A rigorous analysis of these spacetimes would be required to truly prove any predictions that we make.

The metric ansatz imposed a translational symmetry on the torus, allowing only dependence on the $x$-coordinate of the two $(x,y)$ coordinates on the $\mathbb{T}^2$. Future work might explore the consequences of breaking this symmetry, allowing more general perturbations of the spacetime. Similarly, it would be interesting to explore the consequences of choosing a different modular parameter $\tau$ with a nonzero real part for the torus, from both a gravitational and CFT perspective.

Our prediction extended beyond subexponential decay and also included a subleading polynomial time prefactor. We were unable to detect this in our simulations, likely due to its relatively small effect in comparison to the subexponential term. Equally, it may be the case that the prediction of this term is not true, or otherwise incomplete. Uncovering the truth in this story is another area for future exploration.

Lastly, the predicted subexponential decay is only for \textit{analytic} initial data. It would be interesting to explore the decay of data in other regularity classes - for example, data that is not real-analytic but is in some Sobolev space. If the Fourier spectrum of the initial data depends polynomially on the wavenumber rather than exponentially, the predicted late time behaviour is instead a power law.

\acknowledgments
We would like to thank Don~Marolf and Gary~Horowitz for reading an earlier version of this manuscript. JRVC was supported by an STFC studentship ST/W507350/1 and the Cambridge Trust. JES is supported by STFC grant ST/X000664/1 and by Hughes Hall College.

\appendix

\section{Convergence plots}

\begin{figure}[h]
\centering
\begin{tabular}{cc}
     \includegraphics[width=0.435\linewidth]{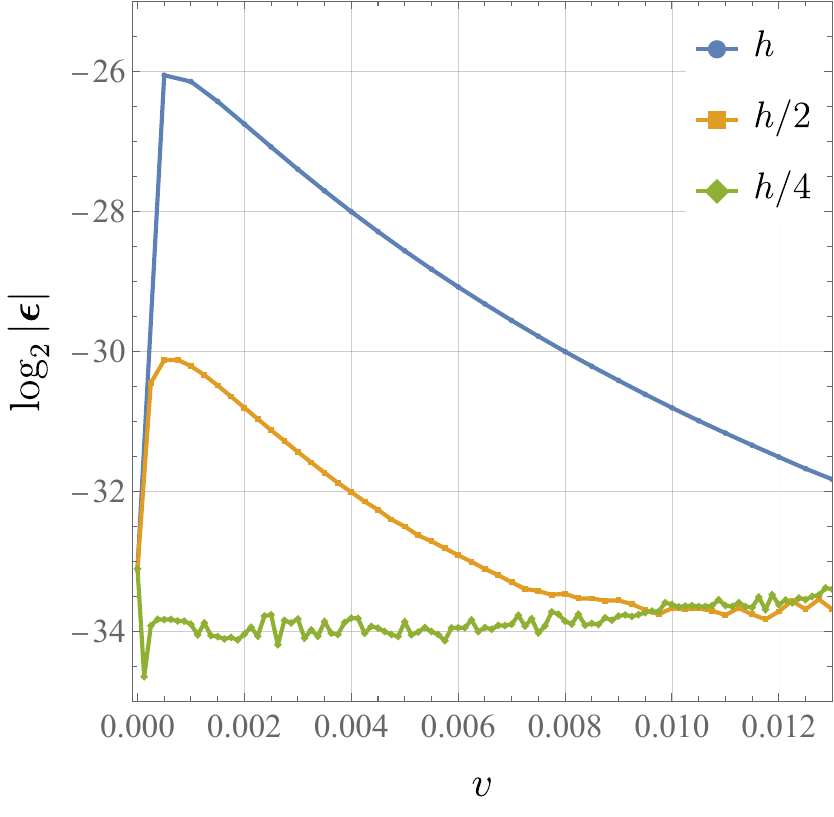} & \includegraphics[width=0.45\linewidth]{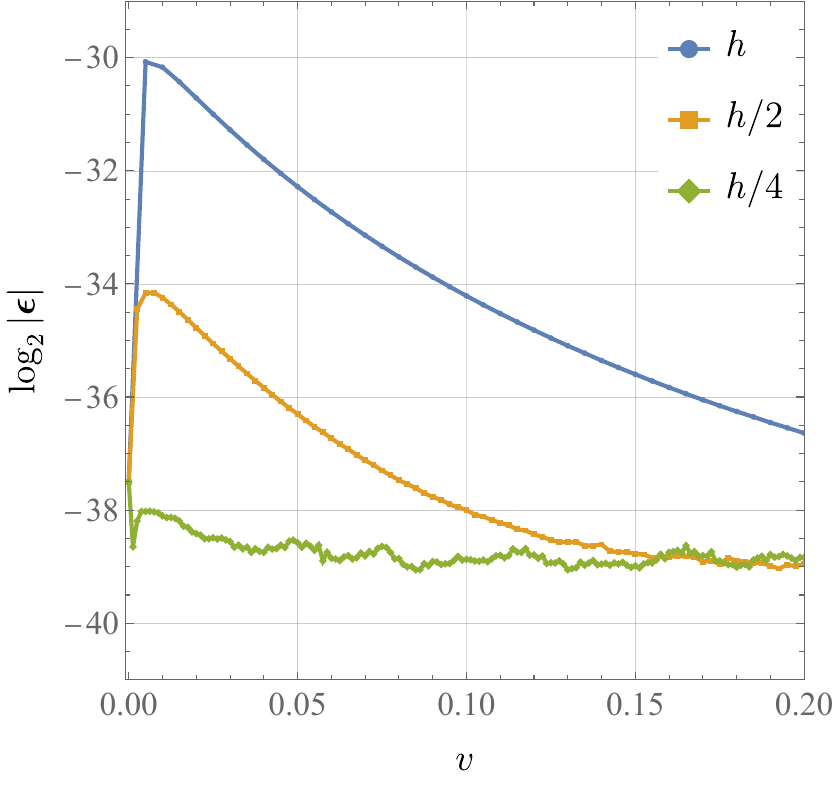}
\end{tabular}
\caption{Convergence plots showing the time dependence of the binary logarithm of the norm of the residual vector. The left plot shows the $L=2\pi/3$ case, and the right plot shows the $L=2\pi$ case. This is shown for evolutions using timesteps $h$, $h/2$ and $h/4$ where $h$ corresponds to the timesteps used in the main text. In both cases, the change in error from the $h$ plot to the $h/2$ plot is consistent with the factor of $1/4$ expected from RK4. The $h/4$ plots have reached the threshold where floating point errors dominate.}\label{fig:convergence_plots}
\end{figure}

To demonstrate the convergence of the RK4 method, we use the apparent horizon condition \eqref{eq:AdS4_horizon_condition} to measure the error as the timestep is decreased. In the marching orders described in \autoref{marching_orders}, we chose to impose the boundary condition for $d_t\chi$ at the AdS boundary $z=0$. This choice corresponds to enforcing the conservation of the boundary stress tensor. An alternative choice of boundary condition would be to impose the apparent horizon condition at $z=1$, written in terms of $d_t\chi$ rather than $\partial_v \chi$. Since both conditions must hold on a true solution, the condition not chosen may be used as a measure of the numerical error. We implement this by defining the residual vector at each timestep $\boldsymbol{\epsilon}$ to be the vector of values of the apparent horizon condition at the grid points. Analysis of RK4 with a timestep $h$ indicates that the error should scale with $h^4$. In \autoref{fig:convergence_plots}, we show the norm of $\boldsymbol{\epsilon}$ for the systems studied in the main text with timesteps $h$, $h/2$ and $h/4$ and observe convergence in agreement with this analysis.

\bibliography{bibliography}{}

@article{Maldacena:1997re,
    author = "Maldacena, J. M.",
    title = "{The Large $N$ limit of superconformal field theories and supergravity}",
    eprint = "hep-th/9711200",
    archivePrefix = "arXiv",
    reportNumber = "HUTP-97-A097, HUTP-98-A097",
    doi = "10.4310/ATMP.1998.v2.n2.a1",
    journal = "Adv. Theor. Math. Phys.",
    volume = "2",
    pages = "231--252",
    year = "1998"
}

@article{Aharony:1999ti,
    author = "Aharony, O. and Gubser, S. S. and Maldacena, J. M. and Ooguri, H. and Oz, Y.",
    title = "{Large N field theories, string theory and gravity}",
    eprint = "hep-th/9905111",
    archivePrefix = "arXiv",
    reportNumber = "CERN-TH-99-122, HUTP-99-A027, LBNL-43113, RU-99-18, UCB-PTH-99-16, LBL-43113",
    doi = "10.1016/S0370-1573(99)00083-6",
    journal = "Phys. Rept.",
    volume = "323",
    pages = "183--386",
    year = "2000"
}

@article{Witten:1998qj,
    author = "Witten, E.",
    title = "{Anti de Sitter space and holography}",
    eprint = "hep-th/9802150",
    archivePrefix = "arXiv",
    reportNumber = "IASSNS-HEP-98-15",
    doi = "10.4310/ATMP.1998.v2.n2.a2",
    journal = "Adv. Theor. Math. Phys.",
    volume = "2",
    pages = "253--291",
    year = "1998"
}

@article{Bhattacharyya:2007vjd,
    author = "Bhattacharyya, S. and Hubeny, V. E. and Minwalla, S. and Rangamani, M.",
    title = "{Nonlinear Fluid Dynamics from Gravity}",
    eprint = "0712.2456",
    archivePrefix = "arXiv",
    primaryClass = "hep-th",
    reportNumber = "TIFR-TH-07-44, DCPT-07-73, NI07097",
    doi = "10.1088/1126-6708/2008/02/045",
    journal = "JHEP",
    volume = "02",
    pages = "045",
    year = "2008"
}

@inproceedings{Hubeny:2011hd,
    author = "Hubeny, V. E. and Minwalla, S. and Rangamani, M.",
    title = "{The fluid/gravity correspondence}",
    booktitle = "{Theoretical Advanced Study Institute in Elementary Particle Physics}: {String theory and its Applications: From meV to the Planck Scale}",
    eprint = "1107.5780",
    archivePrefix = "arXiv",
    primaryClass = "hep-th",
    pages = "348--383",
    year = "2012"
}

@article{Kovtun:2004de,
    author = "Kovtun, P. and Son, Dan T. and Starinets, Andrei O.",
    title = "{Viscosity in strongly interacting quantum field theories from black hole physics}",
    eprint = "hep-th/0405231",
    archivePrefix = "arXiv",
    reportNumber = "INT-PUB-04-09, UW-PT-04-04",
    doi = "10.1103/PhysRevLett.94.111601",
    journal = "Phys. Rev. Lett.",
    volume = "94",
    pages = "111601",
    year = "2005"
}

@article{Policastro:2002se,
    author = "Policastro, G. and Son, D. T. and Starinets, A. O.",
    title = "{From AdS / CFT correspondence to hydrodynamics}",
    eprint = "hep-th/0205052",
    archivePrefix = "arXiv",
    reportNumber = "INT-PUB-02-32",
    doi = "10.1088/1126-6708/2002/09/043",
    journal = "JHEP",
    volume = "09",
    pages = "043",
    year = "2002"
}

@article{Adams:2013vsa,
    author = "Adams, A. and Chesler, P. M. and Liu, H.",
    title = "{Holographic turbulence}",
    eprint = "1307.7267",
    archivePrefix = "arXiv",
    primaryClass = "hep-th",
    reportNumber = "MIT-CTP-4460",
    doi = "10.1103/PhysRevLett.112.151602",
    journal = "Phys. Rev. Lett.",
    volume = "112",
    number = "15",
    pages = "151602",
    year = "2014"
}

@article{Westernacher-Schneider:2017xie,
    author = "Westernacher-Schneider, J. R.",
    title = "{Fractal dimension of turbulent black holes}",
    eprint = "1710.04264",
    archivePrefix = "arXiv",
    primaryClass = "gr-qc",
    doi = "10.1103/PhysRevD.96.104054",
    journal = "Phys. Rev. D",
    volume = "96",
    number = "10",
    pages = "104054",
    year = "2017"
}

@article{Festuccia:2008zx,
    author = "Festuccia, G. and Liu, H.",
    title = "{A Bohr-Sommerfeld quantization formula for quasinormal frequencies of AdS black holes}",
    eprint = "0811.1033",
    archivePrefix = "arXiv",
    primaryClass = "gr-qc",
    reportNumber = "MIT-CTP-3995, SCIPP-08-11",
    doi = "10.1166/asl.2009.1029",
    journal = "Adv. Sci. Lett.",
    volume = "2",
    pages = "221--235",
    year = "2009"
}

@article{Dunn_2016,
doi = {10.1088/0264-9381/33/12/125010},
year = {2016},
publisher = {IOP Publishing},
volume = {33},
number = {12},
pages = {125010},
author = {Dunn, J. and Warnick, C.},
title = "{The Klein–Gordon equation on the toric AdS-Schwarzschild black hole}",
journal = {Classical and Quantum Gravity}
}

@phdthesis{Dunn:2018let,
    author = "Dunn, J.",
    title = "{Aspects of stability of the toroidal AdS Schwarzschild black hole}",
    doi = "10.25560/65669",
    year = "2018"
}

@article{Kehle:2022uvc,
    author = {Kehle, C. and Unger, R.},
    title = {Gravitational collapse to extremal black holes and the third law of black hole thermodynamics},
    journal = {J. Eur. Math. Soc.},
    doi = {10.4171/JEMS/1591},
    year = {2025}
}

@book{Laplace_method,
author = {Bender, C. and Orszag, S.},
year = {1999},
title = "{Advanced Mathematical Methods for Scientists and Engineers: Asymptotic Methods and Perturbation Theory}",
volume = {1},
isbn = {978-1-4419-3187-0},
doi = {10.1007/978-1-4757-3069-2}
}

@article{Balasubramanian_2014,
   title={Losing forward momentum holographically},
   volume={31},
   ISSN={1361-6382},
   DOI={10.1088/0264-9381/31/12/125010},
   number={12},
   journal={Classical and Quantum Gravity},
   publisher={IOP Publishing},
   author={Balasubramanian, K. and Herzog, C. P.},
   year={2014}, pages={125010} }

@article{Crump:2025,
    author = "Crump, J. R. V. and Santos, J. E.",
    title = "{Tails from the Bulk: Gravitational Decay in AdS$_5$}",
    eprint = "2506.18991",
    archivePrefix = "arXiv",
    primaryClass = "hep-th",
    year = "2025"
}

@article{Dunn:2018xdm,
    author = "Dunn, J. and Warnick, C.",
    title = "{Stability of the Toroidal AdS Schwarzschild Solution in the Einstein--Klein-Gordon System}",
    eprint = "1807.04986",
    archivePrefix = "arXiv",
    primaryClass = "gr-qc",
    month = "7",
    year = "2018"
}

@article{Warnick:2013hba,
    author = "Warnick, Claude M.",
    title = "{On quasinormal modes of asymptotically anti-de Sitter black holes}",
    eprint = "1306.5760",
    archivePrefix = "arXiv",
    primaryClass = "gr-qc",
    reportNumber = "ALBERTA-THY-3-13",
    doi = "10.1007/s00220-014-2171-1",
    journal = "Commun. Math. Phys.",
    volume = "333",
    number = "2",
    pages = "959--1035",
    year = "2015"
}

@article{Warnick:2022hnc,
    author = "Warnick, C.",
    title = "{(In)completeness of Quasinormal Modes}",
    doi = "10.5506/APhysPolBSupp.15.1-A13",
    journal = "Acta Phys. Polon. Supp.",
    volume = "15",
    number = "1",
    pages = "1",
    year = "2022"
}

@article{Horowitz:2020tpa,
    author = "Horowitz, G. T. and Wang, D.",
    title = "{Consequences of Analytic Boundary Conditions in AdS}",
    eprint = "2002.10609",
    archivePrefix = "arXiv",
    primaryClass = "gr-qc",
    doi = "10.1007/JHEP04(2020)045",
    journal = "JHEP",
    volume = "04",
    pages = "045",
    year = "2020"
}

@article{McCulloch:2025fzk,
    author = "McCulloch, Ewan and Jacoby, J. Alexander and von Keyserlingk, Curt and Gopalakrishnan, Sarang",
    title = "{Subexponential decay of local correlations from diffusion-limited dephasing}",
    eprint = "2504.05380",
    archivePrefix = "arXiv",
    primaryClass = "quant-ph",
    month = "6",
    year = "2025"
}
\bibliographystyle{JHEP}

\end{document}